\documentclass[
reprint,
superscriptaddress,
 amsmath,amssymb,
 aps,
 prd,
]{revtex4-2}

\usepackage{float}
\makeatletter
\let\newfloat\newfloat@ltx
\makeatother

\usepackage[T1]{fontenc}
\usepackage{graphicx}
\usepackage{dcolumn}
\usepackage{bm}
\usepackage{xcolor}
\usepackage{algorithm}
\usepackage{algpseudocode}
\usepackage[normalem]{ulem}
\usepackage{hyperref}
\usepackage[english]{babel}
\newcommand{\pasp}{{Pub. Astron. Soc. Pac.}}
\DeclareMathOperator*{\argmax}{arg\,max}

\newcommand{\mlk}[1]{\textcolor{red}{(MLK: #1)}}
\newcommand{\nikos}[1]{\textcolor{blue}{(NiKar: #1)}}

\begin{document}

\preprint{APS/123-QED}

\title{An efficient GPU-accelerated multi-source global fit pipeline\\for LISA data analysis}

\author{Michael L. Katz}
\email{michael.l.katz@nasa.gov}
 \affiliation{NASA Marshall Space Flight Center, Huntsville, Alabama 35811, USA}
 \affiliation{Max-Planck-Institut f\"ur Gravitationsphysik, Albert-Einstein-Institut,  Am M\"uhlenberg 1, 14476 Potsdam-Golm, Germany}
\author{Nikolaos Karnesis}%
\affiliation{%
 Department of Physics, Aristotle University of Thessaloniki, Thessaloniki 54124, Greece
}%
\author{Natalia Korsakova}
\affiliation{
 Astroparticule et Cosmologie, CNRS, Universit\'e Paris Cit\'e, F-75013 Paris, France
}%
\author{\\Jonathan R. Gair}
\affiliation{Max-Planck-Institut f\"ur Gravitationsphysik, Albert-Einstein-Institut,  Am M\"uhlenberg 1, 14476 Potsdam-Golm, Germany}

\author{Nikolaos Stergioulas}
\affiliation{%
 Department of Physics, Aristotle University of Thessaloniki, Thessaloniki 54124, Greece
}%

\date{\today}

\begin{abstract}

The large-scale analysis task of deciphering gravitational wave signals in the LISA data stream will be difficult, requiring a large amount of computational resources and extensive development of computational methods. Its high dimensionality, multiple model types, and complicated noise profile require a global fit to all parameters and input models simultaneously. In this work, we detail our global fit algorithm, called ``Erebor,'' designed to accomplish this challenging task. It is capable of analysing current state-of-the-art datasets and then growing into the future as more pieces of the pipeline are completed and added. We describe our pipeline strategy, the algorithmic setup, and the results from our analysis of the LDC2A Sangria dataset, which contains Massive Black Hole Binaries, compact Galactic Binaries, and a parameterized noise spectrum whose parameters are unknown to the user. The Erebor algorithm includes three unique and very useful contributions: GPU acceleration for enhanced computational efficiency; ensemble MCMC sampling with multiple MCMC walkers per temperature for better mixing and parallelized sample creation; and special online updates to reversible-jump (or trans-dimensional) sampling distributions to ensure sampler mixing and accurate initial estimates for detectable sources in the data. We recover posterior distributions for all 15 (6) of the injected MBHBs in the LDC2A training (hidden) dataset. We catalog $\sim12000$ Galactic Binaries ($\sim8000$ as high confidence detections) for both the training and hidden datasets. All of the sources and their posterior distributions are provided in publicly available catalogs.

\end{abstract}

\maketitle


\section{\label{sec:intro}Introduction}

In the mid-2030s, the Laser Interferometer Space Antenna (LISA) will launch into space to measure gravitational waves emanating from a variety of astrophysical and cosmological sources in the millihertz frequency band \cite{LISA:2017pwj, Colpi:2024xhw}. LISA will add important information to the gravitational-wave spectrum, following on the discoveries of ground-based observing networks at higher frequencies \cite{KAGRA:2021vkt} and pulsar timing arrays at lower frequencies \cite{NANOGrav:2023gor, EPTA:2023fyk, Xu:2023wog, Zic:2023gta}. LISA will detect different astrophysical sources: compact object binaries, also referred to as Galactic Binaries (GB), inside and close to the Milky Way Galaxy in slowly evolving orbits \cite[e.g.][]{LISA:2017pwj, LISA:2022yao, Kupfer:2023nqx, Keim:2022lzv, Rieck:2023pej}; Massive Black Hole Binaries (MBHB) in the centers of galaxies \cite[e.g.][]{Begelman:1980vb, LISA:2022yao, Volonteri:2021sfo, Askar:2021rll, Bonetti:2018tpf}; and Extreme Mass Ratio Inspirals where a smaller compact object orbits around a Massive Black Hole \cite[e.g.][]{Babak:2017tow, LISA:2022yao}. LISA will also be able to detect astrophysical and cosmological stochastic signals of sufficient amplitude~\cite[e.g.][]{LISA:2017pwj}. All of the signals will exist simultaneously in the LISA data stream with MBHBs providing quasi-transient signals over days to weeks; GB signals existing for the duration of the mission with almost monochromatic signals; and EMRIs somewhere in between with long-duration ($\sim$year) signals that will merge after a long time inspiralling. 

In addition to this astrophysical variety, LISA will have a complicated noise profile with non-stationary, non-Gaussian noise effects such as sensitivity drift, gaps, and glitches \cite{Baghi:2019eqo, Baghi:2021tfd, Spadaro:2023muy, Digman:2022jmp, Littenberg:2023xpl}. Additionally, the orbit and performance of the LISA experiment will need to be included in any global analysis \cite{Katz:2022yqe, Bayle:2023qfo, Page:2021asu, Page:2023hxm}. There will also be an initial data reduction pipeline that must be run properly prior to performing the scientific data analysis. This step includes instrument calibration~\cite{Savalle:2022xpv} and reduction of primary noises such as laser noise, clock noise, and tilt-to-length (TTL) noise~\cite{Paczkowski:2022nrt, Hartig:2023ofu, Wanner:2024eoa}. The techniques of time delay interferometery~(TDI)~\cite[e.g.][and sources within]{Tinto:2020fcc, Hartwig:2022yqw}, global time synchronization, and ranging~\cite{Reinhardt:2023ccg} will be used to reduce these crucial noise contributions. 

\begin{figure}
    \includegraphics[scale=0.5]{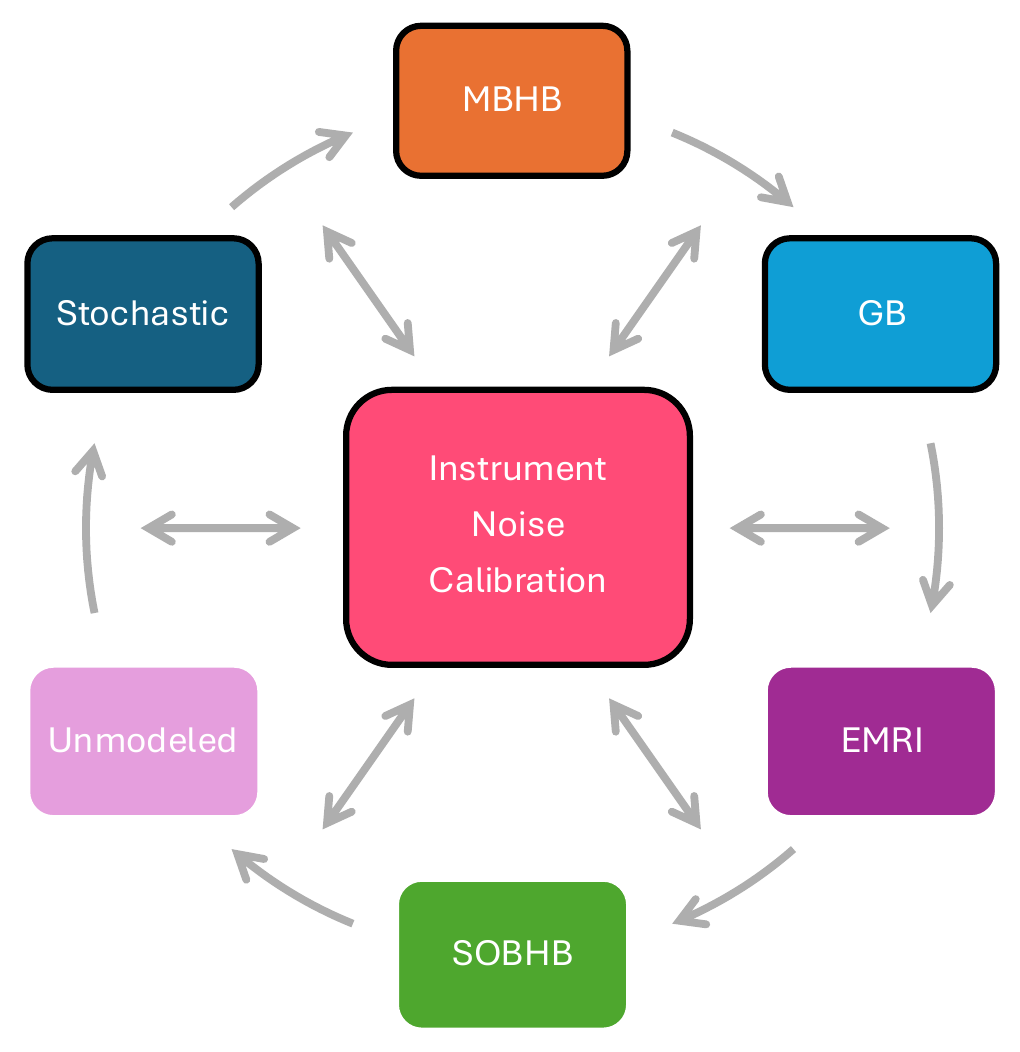}
    \caption{The ``Global Fit Wheel'' representing an example flow for a global fit algorithm across different source and instrument sub-analyses for LISA~\cite{Littenberg:2020bxy}. In clockwise order: massive black hole binaries (MBHB), Galactic binaries (GB), extreme-mass-ratio inspirals (EMRI), stellar-origin black hole binaries (SOBHB), unmodelled gravitational wave sources (e.g. bursts), and stochastic signals like astrophysical and cosmological backgrounds. The instrument, noise, and calibration will enter the process throughout to ensure updated experimental information. Boxes outlined in black show which components are currently included in their \textit{initial} implementation state in Erebor. However, as of now we do not include any instrument or calibration analyses. That is left to future work.}
    \label{fig:wheel}
\end{figure}

This complicated analysis requires a global fit over all parameters characterising the models for the signals and the noise. The design of this type of pipeline has been, and will continue to be, a large developmental 
project requiring many participating groups and expertise. LISA global fit pipelines are likely to be built on the concept of the global fit ``wheel'': separate modules specifically designed for different source or instrumental analyses will run in parallel and communicate with each other throughout the global fit run~\cite{Littenberg:2023xpl}. An example ``wheel'' illustration is shown in Figure~\ref{fig:wheel}. This takes advantage of the small correlations expected between different source classes and instrumental models to perform the complex analysis in a much more computationally friendly manner. 

In order to benchmark and test the implementation of these algorithms across the LISA community, the LISA Data Challenges Working Group (LDCWG) has released a series of datasets of simulated LISA data. The LDC2A dataset, also called ``Sangria''~\cite{sangria}, was formulated to test the current state of the art in the LISA global fitting community: simultaneous fitting of MBHBs, GBs throughout the Galaxy, and a stationary, Gaussian and overall simplified noise profile. It contains one year of data with the aforementioned signals and noise combined into a single dataset. It also comes with information related to the exact input populations to be used for debugging and developing experimental pipelines \cite{ldc}.

There are a few groups working on this large-scale problem. In \cite{Littenberg:2020bxy, Littenberg:2023xpl, Lackeos:2023eub}, the authors describe a pipeline that can successfully analyse the LDC2A dataset and produce output posterior distributions for the detected sources and noise properties. They also provide details of their methods and findings and discuss future projects related to their global fit pipeline.

Recently, \cite{Strub:2024kbe} demonstrated a highly-efficient maximum Likelihood solution for GBs and MBHBs, which also included a fit to the overall noise profile including the confusion foreground signal generated by the many unresolvable GBs. They performed this analysis for one year of data, updating week by week, thus showing its utility and capabilities to handle time-evolution. This useful pipeline will provide a low-cost initial estimate of source parameters, populations and noise properties. The information contained in this estimate can also be used to further improve the efficiency of full posterior estimation-based global fit samplers.

In this work, we present our implementation of a LISA global fit pipeline for analyzing the sources and noise profile in the LDC2A dataset. We perform this full computation on the ``training'' data and, for the first time, on the ``hidden'' dataset provided by the LDC. Our goal was to learn valuable lessons from \cite{Littenberg:2023xpl}, while trying to build an independent algorithm with important differences in order to further expand the LISA global fit knowledge space. 

Our pipeline follows a similar Bayesian approach as presented in \cite{Littenberg:2023xpl}. We use Reversible-Jump Markov Chain Monte Carlo (RJMCMC; also known as Trans-Dimensional Markov Chain Monte Carlo) techniques \cite{rjmcmc1} to fit for and characterize the uncertain number of GB sources in the data. We also adapt the concept of the global fit ``wheel'' (Figure~\ref{fig:wheel}) running large modules of the sampler in a blocked Gibbs fashion \cite{gibbs1, gibbs2, gibbs3}, sending out and receiving updates to and from all other modules periodically \cite{Littenberg:2023xpl}.

There are three important large-scale differences between our pipeline and \cite{Littenberg:2023xpl}. The pipeline in \cite{Littenberg:2023xpl} used a large number of CPUs run in parallel. In contrast, we designed our algorithm specifically to maximize the efficiency and usage of Graphics Processing Units (GPU). GPUs are specialized hardware that are designed for a high level of parallel processing. GPUs are generally more cost and energy efficient than CPU clusters with equivalent computational power and are expected to continue to improve at a more rapid rate than CPUs \cite{gpu_better_efficiciency}. For these reasons, we designed our algorithm for GPUs in order to take advantage of the computational and energy efficiency, which we expect to further improve over the coming years. 

The second main difference is we use an Ensemble Sampling setup for our Markov Chain Monte Carlo runs \cite{emcee-citation}. Rather than having one walker per temperature as used in \cite{Littenberg:2023xpl}, we use many walkers per temperature by leveraging the MCMC software package \texttt{eryn} \cite{Karnesis:2023ras, eryn_zenodo, emcee-citation}. A ``walker'' is one point in parameter space representing a current state of the sampler. As walkers evolve due to probabilistic distributions, each state they evolve through forms the chain of samples that comes together to make the posterior distribution. In addition to the intrinsic benefits of ensemble sampling, within the LISA global fit context it allows us to marginalize over the various model types more efficiently: walkers can contain different residuals or noise estimates that come from the random differences during sampling of other parts of the pipeline. Ensemble sampling is also better for parallelization of MCMC sample creation: there is more than one walker in the cold-chain; each walker can be evaluated independently in parallel across the GPU. We also employ the ``group stretch'' MCMC proposal, which requires an ensemble setup to use and minimal tuning compared to typical covariance proposals~\cite{Karnesis:2023ras}.

Finally, using one or more extra GPUs, we perform online search and refitting techniques to continue to improve and update our RJMCMC proposals as the sampling run proceeds. This allows our algorithm to adapt and ensure all detectable sources are found and characterized, as well as to maintain strong mixing in the sampler. This also makes it much easier for us to restart our global fit at later times rather than requiring to build it up over time. 

Our pipeline finds all 15 (6) MBHB sources in the training (hidden) data and provides posterior estimates for all of them. We determine the posterior distributions on parameterized models of the detector noise and galactic foreground noise components. We produce a catalog of $\sim12000$ GB sources ($\sim8000$ with high confidence) and characterize the uncertainty in the template (or source) count found in the sampler at various times. 

In the next section (Section~\ref{sec:sources}), we will detail and explain the gravitational-wave signals and Likelihood computations for MBHB and GB signals. In Section~\ref{sec:noise}, we explain our parameterized model for the fit to the noise and foreground confusion. An overview of the MCMC methods employed is provided in Section~\ref{sec:mcmc}. In Section~\ref{sec:pe}, we detail our full global fit algorithm. We follow that with our results for our LDC2A analysis in Section~\ref{sec:res} and a broader discussion of those results in Section~\ref{sec:disc}. We end with some concluding remarks in Section~\ref{sec:conclusion}.

\section{\label{sec:sources}Gravitational-wave Likelihoods}

For the current implementation of the pipeline, there are two gravitational-wave sources: Massive Black Hole Binaries (MBHB) and Galactic Binaries (GBs). These are the two types of sources included in the LDC2A dataset (also called ``Sangria''). This dataset is the first LDC dataset designed to test multiple-source pipelines.

The two sources have vastly different waveform properties and generation methods. Here, we will detail the overall properties that apply to both and describe the specific models used to generate the signal templates. 

Gravitational-wave signals are real-valued signals in the time domain denoted by $s(t)$. These signals are measured by the detector by observing their projection on the sensitive axes of the detector. In LISA, multiple signals will be detected simultaneously: $S(t)=\sum_j\sum_{i\in j}s_{ij}$ with $j$ indicating the source type (MBHBs or GBs) and $i$ indexing the sources within type $j$. The output of the detector measurements, from a data analysis perspective, are these signals combined with the noise in the detector: $d(t) = S(t) + n(t)$, where $n(t)$ is the noise component. To match these signals, we construct template waveforms, $H(t)=\sum_j\sum_{i\in j}h_{ij}$, built from theory and use these to detect and statistically characterize the signals present in the data stream. We assume that the noise is distributed according to a Gaussian distribution; therefore, we use a Gaussian Likelihood based on the residual between the data stream and the specified combination of template waveforms. We will discuss more about this choice of Likelihood and the noise properties in Section~\ref{sec:noise}. The natural log of the Likelihood is given by,
\begin{equation}\label{eq:like}
    \ln{\mathcal{L}} = -\frac{1}{2}\langle d - H | d - H\rangle - \sum_f \ln{S_n(f)},
\end{equation}
where $\sum_f\ln{S_n(f)}$ needs to be included when there is noise uncertainty, with $S_n(f)$ representing the Power Spectral Density of the noise. We will also refer to this Likelihood as the ``base Likelihood`` or $\mathcal{L}_b$. The first term on the right hand side is an inner product:
\begin{equation}\label{eq:inner}
    \langle b(t)|c(t) \rangle = 4 \text{Re} \sum_f\frac{\tilde{b}(f)^*\tilde{c}(f)}{S_n(f)}\Delta f,
\end{equation}
for real-valued time-series $b(t)$ and $c(t)$, where $\tilde{b}(f)$ denotes the Fourier Transform of $b(t)$
and $\Delta f$ is the frequency spacing of the Discrete Fourier Transform. This term quantifies the quality of the template fitting by looking at the size of the residual. Both summations in Equations~\ref{eq:like} and~\ref{eq:inner} are over all \textit{positive} frequencies contained in the Discrete Fourier Transform of the observed discretely-sampled time-domain signals.

There are a few useful quantities that we will also define here 
that will be helpful later on. When there is \textit{only} one source in the data $H(t)\equiv h_1(t)$. In these circumstances, it can be useful to separate the source term in Equation~\ref{eq:like} into constituent pieces using the bilinearity of the inner product:
\begin{equation}
    \ln{\mathcal{L}}_1 \propto -\frac{1}{2} \left(\langle d|d \rangle + \langle h_1|h_1 \rangle - 2\langle d|h_1 \rangle \right).
\end{equation}
The optimal signal-to-noise ratio~(SNR) of the template waveform is then given by $\sqrt{\langle h_1|h_1 \rangle}$. The observed SNR of the source (when it is the only source in the data) is $\langle d|h_1 \rangle / \sqrt{\langle h_1|h_1 \rangle}$. When there are multiple sources in the data, the observed SNR of a single source is more complicated to define because of contributions from overlaps with the other sources in the data. 

For the purposes of this work, we use two ``special'' Likelihood formulations that are related to adjusting the data stream by one source. The first is the change in log-Likelihood associated with adding a single source, $a(t)$, to the current set of templates ($H + a$):
\begin{align}\label{eq:addsource}
\begin{split}
    \ln{\mathcal{L}}_{+1} &\propto -\frac{1}{2}\langle d - (H + a) | d - (H + a)\rangle\quad , \\
    &= \ln{\mathcal{L}_b} -\frac{1}{2}\left[\langle a|a \rangle - 2\langle d - H|a \rangle\right]\ , \\
    \therefore\ \Delta \ln{\mathcal{L}}_{+1} &= -\frac{1}{2}\left[\langle a|a \rangle - 2\langle d - H|a \rangle\right].
\end{split}
\end{align}
In other words, if we have already calculated the base Likelihood (Equation~\ref{eq:like}), we can determine the Likelihood after adding one source by only computing terms related to the addition of the single template. This helps in the frequency domain to restrict the summation in Equation~\ref{eq:inner} to frequencies at which the new signal is present. This is especially useful for individual GBs as they have a very small range of frequencies over which they emit gravitational radiation. Similarly to Equation~\ref{eq:addsource}, we can also write down the change in Likelihood due to the swapping of one source in the current template, $r(t)$, for another, $a(t)$:
\begin{align}\label{eq:swapsource}
\begin{split}
    \ln{\mathcal{L}}_{+1/-1} &\propto -\frac{1}{2}\langle d - (H + a - r) | d - (H + a - r)\rangle\quad ,  \\
    &= \ln{\mathcal{L}_b} -\frac{1}{2}\left[\langle a|a \rangle+  \langle r|r \rangle - 2\langle a|r \rangle\right. \\ &\left.\qquad\qquad - 2\langle d - H|a \rangle + 2\langle d - H|r \rangle\right]\ \ , \\
    \therefore\ \Delta \ln{\mathcal{L}}_{+1/-1} &= -\frac{1}{2}\left[\langle a|a \rangle+  \langle r|r \rangle - 2\langle a|r \rangle\right. \\ &\left.\qquad\qquad - 2\langle d - H|a \rangle + 2\langle d - H|r \rangle\right]\ .
\end{split}
\end{align}
This once again allows for single-source-width computations. However, in Equation~\ref{eq:swapsource}, the number of frequencies needed to evaluate the $\langle a|r \rangle$ term will be between the frequency width of one and two sources depending on how much $a(t)$ and $r(t)$ overlap in frequency. If they perfectly overlap, then the effective width is the width of one source. If they do not overlap at all, the effective width is the combination of the frequency widths of both sources. If they overlap imperfectly, the width will be between these two extremes. 

As previously mentioned, observations from the triangular LISA detector will be pre-processed on-ground to reduce laser frequency noise using TDI~\cite{Tinto:2020fcc}. The outputs of this pipeline are called ``TDI observables''. The most basic set of these are the observables constructed from each set of two arms: $X_1,Y_1,Z_1$. The ``1'' denotes the \textit{1.5} Generation TDI observables~\cite{Tinto:1999yr, sangria_doc}. The Sangria dataset uses the 1.5 Generation TDI observables, and assumes equal and constant detector arm lengths (TDI Generation 1.5 can handle a rigid and rotating LISA constellation; the armlengths do not have to be equal). During the actual mission, we will be using Second Generation TDI observables~\cite{Shaddock:2003dj}, which are slightly more complicated. These observables will account for the rotation and flexing of the LISA constellation, meaning the armlengths can vary with time. This is an ongoing area of study and will be added into the pipeline presented here as a part of future work. 

The $X_1,Y_1,Z_1$ variables are correlated in their noise properties. Therefore, we make a linear transformation to $A_1,E_1,T_1$ variables~\cite{Prince:2002hp, Vallisneri2005}, which, under idealised assumptions about the LISA instrument, would have uncorrelated noise properties:
\begin{subequations}
\begin{align}
    A_1 =& \frac{1}{\sqrt{2}}\left(Z_1-X_1\right)\ , \\
    E_1 =& \frac{1}{\sqrt{6}}\left(X_1-2Y_1+Z_1\right)\ , \\
    T_1 =&\frac{1}{\sqrt{3}}\left(X_1+Y_1+Z_1\right)\ .
\end{align}
\end{subequations}
For the sake of simplicity, in this work we do not consider the $T_1$ channel as it generally suppresses gravitational-wave signals. These uncorrelated variables simplify the analysis, but, as in the case of First versus Second Generation TDI, they will not be used in the final analysis. A more complicated Likelihood matrix will be needed to properly take into account the existence of correlations between the channels. 

\subsection{Input Template Models}

All waveform models used in this pipeline as templates are mathematically equivalent to those used in previous works. For that reason, we are not going to describe all of the details. Rather, we point the reader to the necessary citations to understand the specifics of the template generation methods used.

\begin{figure}
    \centering
    \includegraphics[width=\linewidth]{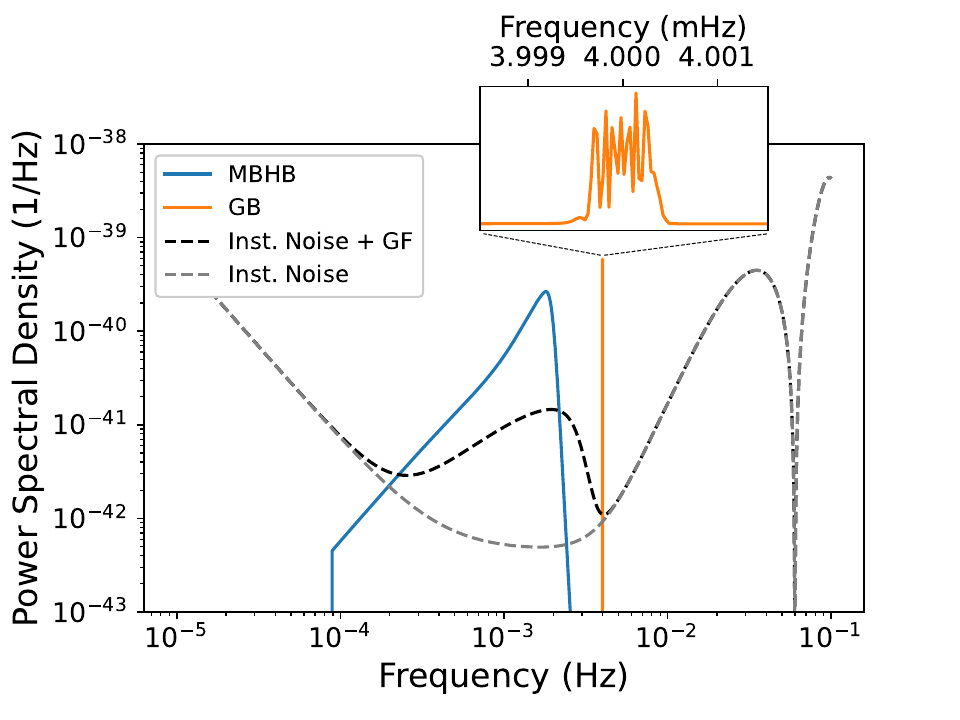}
    \caption{An example of an MBHB waveform, GB waveform, LISA instrumental noise, and the Galactic foreground (GF) confusion contribution. The MBHB waveform is shown in blue. The GB waveform is shown in orange. The GB waveform is very narrow-band in frequency. For this reason, we also show a zoom-in on the GB signal.}
    \label{fig:initial_ex}
\end{figure}

The MBHB waveforms are those used in~\cite{Katz:2021uax}. One example waveform for an MBHB is shown in the frequency domain in Figure~\ref{fig:initial_ex}. We use the aligned-spin IMRPhenomD waveform model~\cite{Khan2016, Husa2016}. A few computational improvements have been made since that paper. The most important of these is a combination of the waveform generation with the Likelihood computation. Previously the waveform was constructed and then stored in memory. It would then be passed to a Likelihood function where the Likelihood would be computed. Now, as each value of the waveform is generated at each frequency, within the same GPU kernel the Likelihood at that frequency is immediately computed and entered into the full Likelihood summation. This avoids the need to store full waveforms in memory, allowing us to run more Likelihood computations simultaneously across a single GPU.

The GB waveforms are based on the FastGB~\cite{Cornish:2007if} algorithm. FastGB is an efficient GB waveform generation method that works by leveraging the small $\dot{f}$ observed in GB sources to separate the waveform generation into a fast time scale piece and a slow time scale piece. They were updated and used for circumbinary exoplanet analysis with LISA in \cite{Katz:2022izt}. For this work, we re-implemented FastGB in a special form for efficient usage on the GPU. This is discussed further in Section~\ref{sec:gb_waveform} in the Appendix. An example GB waveform is shown in Figure~\ref{fig:initial_ex}, as well as a zoom-in on its very localized frequency domain structure.

For more specific information on each waveform model, we refer the reader to the publications cited above.

\section{\label{sec:noise}Noise Treatment}

In the Sangria dataset, as well as our analysis, the detector noise is modelled as stationary and Gaussian. In reality we expect a variety of non-stationary effects including data gaps, glitches, and sensitivity drift over time~\cite{Colpi:2024xhw}. The effect of removing these assumptions is a topic of ongoing research and other LDC datasets will be produced for this purpose. In the future, these advances will be combined with the multiple-source pipeline presented here as those techniques are further developed. 

Under the assumption of stationarity and Gaussianity, the noise covariance matrix in the frequency domain reduces to a vector (the matrix diagonal) due to the absence of correlations between different frequency components. Hence, the summations shown in Equations~\ref{eq:like} and~\ref{eq:inner} are one-dimensional over frequency rather than two-dimensional, which would be needed if there were off-diagonal elements of the noise covariance matrix. 

The exact noise in the detector is uncertain and needs to be fitted as an additional component of the analysis. There are a few different suggested methods for this process. We consider a spectral model for the noise as the sum of two major noise components: the Acceleration and Optical Metrology System (OMS) noises. The overall PSD, $S_n$, in the $A$ and $E$ channels is given by~\cite{QuangNam:2023dgm},
\begin{align}\label{eq:psd}
\begin{split}
    S_n(x) = 8\sin^2{x} \left[2S_\text{pm}(3 + 2\cos{x} + \cos{2x})\right. \\ \left. + S_\text{op}(2+\cos{x})\right]\ ,
\end{split}
\end{align}
where $x=2\pi (f / f^*)$ with $f^* = c / L$ (the speed of light divided by the LISA armlength). $S_\text{pm}$ is related to the acceleration noise, $S_\text{acc}$, by,
\begin{align}
\begin{split}
    S_\text{pm}(f) = &S_\text{acc} \left(2\pi c f \right)^{-2}\\
    \times &\left[1 + \left(\frac{0.4\times10^{-3}~\text{Hz}}{f}\right)^2\right] \left[1 + \left(\frac{f}{8\times10^{-3}~\text{Hz}}\right)^4\right].
\end{split}
\end{align}
Similarly, $S_\text{op}$ is related to the OMS noise, $S_\text{oms}$:
\begin{equation}
    S_\text{op}(f) = S_\text{oms} \left(\frac{2\pi f}{c}\right)^2 \times \left[1 + \left(\frac{2\times10^{-3}~\text{Hz}}{f}\right)^4\right].
\end{equation}
In our global fit pipeline, we fit for $\sqrt{S_\text{oms}}$ and $\sqrt{S_\text{acc}}$ in both channels (i.e., four parameters in total). The injection noise for the LDC2A was generated using \mbox{$\sqrt{S_\text{oms}}=7.9\times10^{-12}$m Hz$^{-0.5}$} and \mbox{$\sqrt{S_\text{acc}}=2.4\times10^{-15}$ m s$^{-2}$ Hz$^{-0.5}$}~\cite{sangria_doc}. Figure~\ref{fig:initial_ex} shows the injection noise PSD as the grey dashed line.

Please note the above model is used for fitting the data. Due to a previous omission from the LDC documentation~\cite{ldc}, we did not include a third noise, backlink noise, that \textit{was} included in the LDC2A dataset noise injection. We discuss this further in the Section~\ref{sec:res}.


\subsection{Galactic foreground noise parameterization}\label{sec:fg}

To characterize the Galactic foreground noise we also choose a simple parameterized model that describes the shape of the expected foreground spectrum. In future work, we will be exploring other potential models to parameterize  the noise and foreground.

Our foreground noise model contains five parameters: $A_{fg}$ represents the overall amplitude of the foreground; $\alpha$ is a power law index; $f_k$ is the ``knee'' frequency where the underlying slopes change; and $s_1$ and $s_2$ are the slopes before and after the knee frequency, respectively. The analytic formula for the foreground noise \textit{sensitivity}, $h_{fg}$, is given by \cite{PhysRevD.104.043019}, 
\begin{align}
\begin{split}
    h_{fg}(f) = &\frac{1}{2}A_{fg}f^{-7/3}e^{-s_1\left(f^\alpha\right)} \\
    &\times \left[1+\tanh{\left(-s_2\left(f - f_k\right)\right)}\right].
\end{split}
\end{align}
The contribution to the PSD in the A and E channels from the galactic foreground is given by,
\begin{equation}
    S_{fg} = 6x^2\left(\sin^2{x}\right)h_{fg}(f)\ \ ,
\end{equation}
where $x=2\pi (f / f^*)$ as in the previous section. This contribution is directly added to $S_n$ given in Equation~\ref{eq:psd} to give the total PSD used in the inner product (Equation~\ref{eq:inner}). Please note that, while we parameterize the instrument noise separately for the A and E channels, we parameterize the foreground contribution as one foreground of five parameters that is then shared across both channels.

An example foreground is shown in Figure~\ref{fig:initial_ex} for parameters $(A_{fg}, \alpha, f_k, s_1, s_2)\approx(3.3\times10^{-44}, 1.2, 2.8\text{ mHz}, 1763.3, 1686.3)$.

\section{Markov Chain Monte Carlo}\label{sec:mcmc}

Throughout the entire pipeline, both in search and parameter estimation, Markov Chain Monte Carlo (MCMC) techniques are heavily used. Here we will give a quick primer and references for these MCMC methods. As we describe further details of the full pipeline later, we will indicate where each of these methods is used. We begin with Bayes' rule:
\begin{equation}\label{eq:bayes}
    p(\vec{\theta}|D, \Lambda) = \frac{p(D|\vec{\theta}, \Lambda)p(\vec{\theta})}{p(D|\Lambda)} = \frac{\mathcal{L}(\vec{\theta})p(\vec{\theta}|\Lambda)}{Z(\Lambda)},
\end{equation}
where $p(\vec{\theta}|\Lambda)$ is the prior probability for model $\Lambda$; $p(\vec{\theta}|D, \Lambda)\equiv \pi(\vec{\theta})$ is the posterior probability of the parameters of model $\Lambda$ given data $D$; $p(D|\vec{\theta}, \Lambda)\equiv \mathcal{L}(\vec{\theta})$ is the Likelihood function; and \mbox{$p(D|\Lambda)\equiv Z(\Lambda)=\int_{\vec{\theta}}\mathcal{L}(\vec{\theta})p(\vec{\theta})d\vec{\theta}$} is the evidence for the model $\Lambda$.

We build our entire pipeline around the \texttt{eryn} sampler \cite{Karnesis:2023ras, eryn_zenodo, emcee-citation}, which is an ensemble-based sampler (ensemble: multiple walkers per temperature rather than just one) designed for use in both fixed-dimensional and trans-dimensional MCMC settings.

As discussed in previous global fit work \cite{Littenberg:2023xpl}, the number of templates needed to fit the data is uncertain. Therefore, we have to use sampling methods that allow for the addition and removal of templates over the course of an MCMC run. For this, we employ the use of Reversible-Jump (or trans-dimensional) MCMC \cite{rjmcmc1}. ``Between-model'' proposals are so-called because they propose to change the underlying model or model dimension ($\Lambda$) rather than just the model parameters ($\vec{\theta}$).

In our situation, where all templates are of the same kind, the addition and subtraction of templates is considered ``nested,'' which vastly simplifies our problem. In a nested setting, this means changing the source count by +1 or -1. The acceptance fraction for nested models when adding one component is given by \cite{nested_models},
\begin{equation}\label{eq:rj-acceptance}
    \alpha_\text{RJ} = \text{min}\left[ 1,  \frac{\mathcal{L}(\vec{\theta}_{k,+1})p(\vec{\theta}_{+1})}{\mathcal{L}(\vec{\theta}_{k})q(\vec{\theta}_{+1})}\right]\ \ ,
\end{equation}
where $k$ is the current number of templates in the data, ``+1'' indicates the proposed model addition, and $q$ is the proposal distribution. In other words, the acceptance fraction becomes the ratio of the Likelihood values of the new and previous point multiplied by the ratio of the prior density to the proposal density of the newly added source. The prior terms for the first $k$ binaries, $p(\vec{\theta}_k)$, cancel out because they are found in both the numerator and the denominator. When proposing to remove a source, the acceptance fraction is inverted and the ``+1'' is replaced with a ``-1'' now indicating the source that is to be removed. At the moment we only use RJMCMC in the GB sampler. All RJ moves within that sampler are based on the acceptance ratio given in Equation~\ref{eq:rj-acceptance}, but they have different proposal distributions ($q(\vec{\theta}_{+1})$). These distributions will be described below.

MCMC moves that conserve the model type and/or the model dimensionality (or source count) are referred to as ``in-model'' or ``within-model'' proposals. These proposals only change the parameters contained in the current model. In \textit{fixed-dimensionality or fixed-model} settings where Reversible Jump is not needed, we use proposals built around the affine-invariant or ``stretch'' proposal \cite{affine1}. The stretch proposal relies on the ensemble setup where multiple MCMC walkers are run simultaneously. As a reminder, a walker represents an instance of a moving sampler state in parameter space that will build up the MCMC chain samples over time. The stretch proposal moves a given walker using the parameter space locations of the other walkers in the ensemble:
\begin{equation}\label{eq:stretch}
    Y_k = X_j + z\left( X_k - X_j \right),
\end{equation} 
where $X_k$ and $Y_k$ are the current and proposed positions of the walker of interest, respectively; $X_j$ is another walker drawn uniformly from the remaining members of the ensemble (excluding the $k$th walker of interest); and $z$ is a randomly distributed variable drawn from a distribution with specific requirements~\cite{affine1}. A main benefit of this type of proposal is that it does not require specific tuning-based adjustments. For more information on the stretch proposal see~\cite{affine1}. In the PSD and MBHB portions of the global fit sampler we use the regular stretch proposal as described in \cite{emcee-citation}.

In our GB sampler where we employ Reversible Jump, we use the ``group stretch'' proposal discussed in \cite{Karnesis:2023ras} because the regular stretch proposal does not scale properly to a Reversible Jump setting where there is, in general, a different number of model instances for each walker. The key difference between the base stretch and group stretch proposals is the determination of $X_j$ in Equation~\ref{eq:stretch}. Rather than drawing from the current walkers of the ensemble, $X_j$ is drawn from a ``stationary group''~\cite{Karnesis:2023ras} that stays fixed through a large number of proposals (required to satisfy detailed balance). For this group, we use a flattened set of all templates contained in the cold-chain in all current walkers within the sampler. Individual groups are formed by sorting this large set of templates by frequency and taking the $n_\text{friends}$ closest sources in frequency to the GB of interest.  We update this group every $\sim50$ proposals and are careful to satisfy detailed balance during the change over. The purpose of this setup is to facilitate efficient in-model sampling in RJMCMC without needing to calculate or tune a covariance matrix or similar such proposal.

We employ the technique of parallel tempering in all modules of our pipeline. We discuss the specifics of our tempering implementation in Section~\ref{sec:parallel-temp} in the Appendix. There, we include details about how a typical tempering scheme must be adjusted to handle a residual-based analysis as is used here.

\section{\label{sec:pe}The Global Fit}

\begin{figure*}
\includegraphics[scale=0.6]{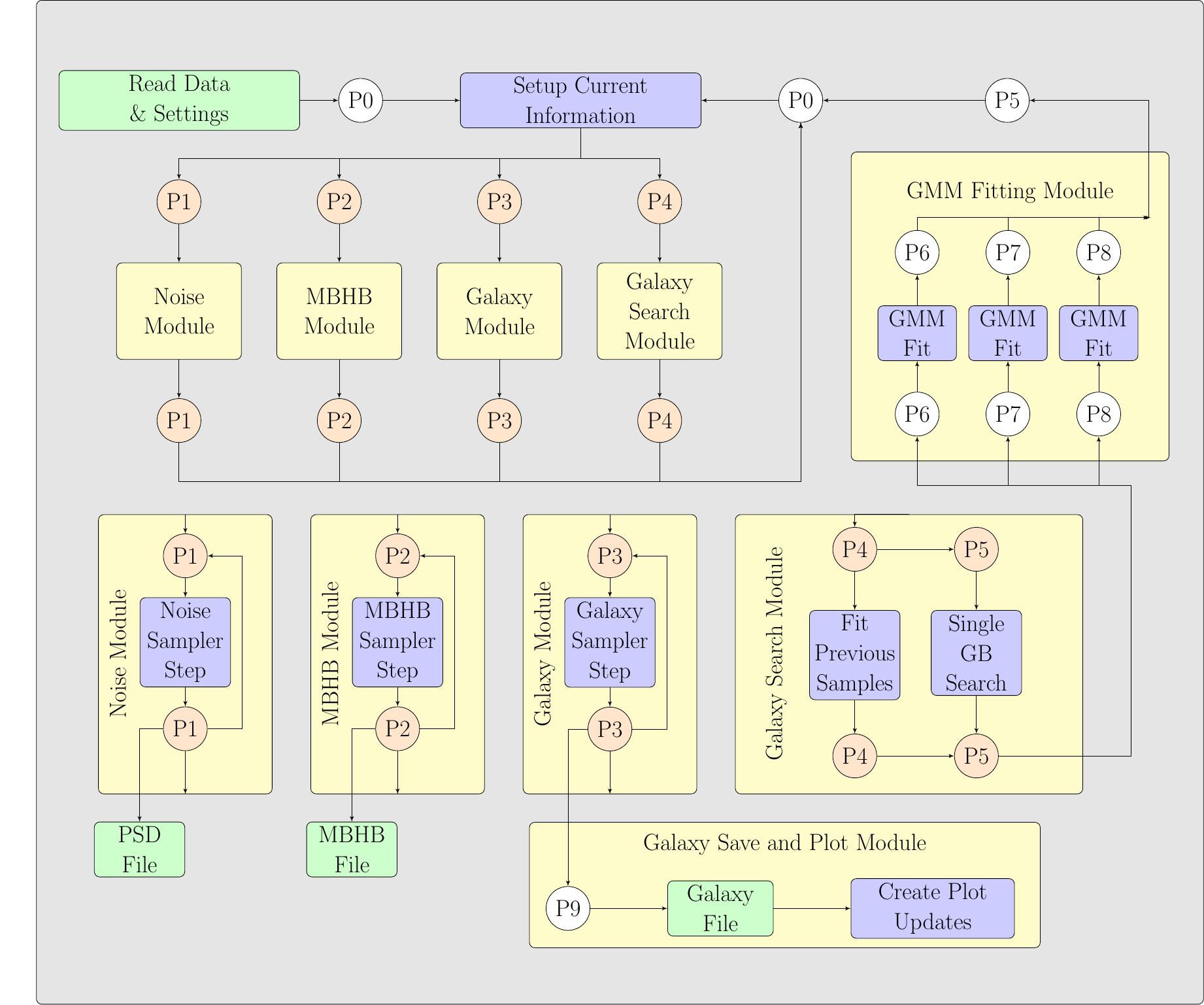}
\caption{Process diagram for the Erebor algorithm. Each circle with ``P\#'' indicates a separate process. Orange circles indicate processes that communicate with and use GPUs. P1, P2, and P3 communicate with one GPU each. P4 and P5 communicate with one GPU total between the two. File-related operations are in green. Separate but communicating modules are shown in yellow. Major algorithmic partitions are given in blue.}
\label{fig:process_diagram}
\end{figure*}

The global fit algorithm presented in this work is very similar in nature to the methods presented in \cite{Littenberg:2023xpl}, with some important differences. The main idea is that the sampler uses a Blocked Gibbs sampling technique \cite{gibbs1, gibbs2, gibbs3} to analyze the MBHBs, PSD (and foreground noise), and GBs as separate sampler entities. Then, periodically, each individual sampler will read out its current cold-chain information to a controlling process. The controlling process will then transmit this to all participating individual samplers so that they update their current residuals and PSD information. It is this operation that marginalizes over other source and PSD posterior distributions contained in the larger global fit. This is effectively the ``wheel'' method shown in Figure~\ref{fig:wheel}.

Our global fit algorithm has four simultaneously running modules: PSD and foreground fitting; MBHB fitting; GB fitting; and a fourth module that contains a search and updating proposal setup for the GBs. A schematic diagram of this process is shown in Figure~\ref{fig:process_diagram}. 

\subsection{Noise and stochastic  foreground signal PSD fitting}\label{sec:pe_psd}

The PSD and foreground noise fitting module is the most straightforward of the four. As described above, each channel (A and E) will have a set of two parameters (acceleration noise and OMS noise) to describe the base LISA sensitivity. This results in four total parameters representing the PSD, with the two sets of parameters acting only on their associated channel.


The foreground noise, as detailed in Section~\ref{sec:noise}, is parameterized by five parameters (amplitude, power-law index, knee frequency, and slopes before and after the knee frequency). In contrast to the four PSD parameters, the foreground noise represents an astrophysical model meaning it is computed independently of each channel and is then transformed and added across both A and E channels.

MCMC sampling of these parameters directly uses \texttt{eryn} and its constituent resources. A fixed-dimensional MCMC is run with ten temperatures ranging from the cold chain sampling the posterior ($T=1$) to an infinite temperature chain sampling the prior distribution, with eight geometrically spaced temperature rungs in between that are allowed to adapt over time~\cite{Vousden2016}. The regular stretch proposal is used in this fixed-dimensional application.

Before beginning to search for astrophysical signals, we run the PSD and foreground sampling operation until it converges to an initial estimate. This estimate is usually okay for the instrumental noise parameters; however, the foreground fit strongly compensates for the lack of subtraction of GW signals. Over the course of the global fit run, as GW signals are found and removed from the residual, the foreground and instrumental noise parameters converge to a their proper state. 

During each update from the controlling process, residuals from the GB and MBHB modules are drawn uniformly from the batch of cold-chain residuals coming in. Each randomly drawn residual is assigned to one walker in the PSD cold chain. Then, in a vertical fashion, the residual assigned to walker $w$ in the cold chain is used for all walkers of index $w$ in all of the higher temperature chains. With this in mind, and for computational efficiency, the ensemble permutations during temperature swapping, which is the default in \texttt{eryn}, is turned off. Temperature swaps are, therefore, only proposed for walkers with the same walker index $w$ located in neighboring temperature chains. With this setup, it is not required to calculate new Likelihoods as described in Section~\ref{sec:parallel-temp} of the Appendix because the underlying residual and PSD against which the Likelihoods are computed are the same.

\subsection{Searching for MBHBs}

With the initial noise PSD fitting complete, we extract the maximum Likelihood PSD and foreground noise components. As previously mentioned, the foreground noise at this point is much higher than it should be, since no astrophysical signals have been subtracted. The MBHBs in the LDC2A dataset are loud enough that all sources can still be detected in this setting. In the future, when lower SNR MBHB signals are included, this method will need to be extended as the confusion noise component is successively reduced. We will discuss this further in Section~\ref{sec:disc}.

The 1-year dataset is split into multiple segments, usually determined by the number of GPUs available. In our case, running four GPUs, we split the dataset into eight segments, running two of them per GPU so that we are analyzing the full year simultaneously. 

Within each segment, we further split the larger pieces into chunks of 3/2 days, overlapping the first 1/2 day and last 1/2 day with neighboring chunks. Therefore, each chunk focuses on a specific 1/2 day, but it contains data from the 1/2 day before and 1/2 day after. The purpose of this is to remove, as best as possible, any confusion between MBHB mergers, which can inhibit the search.

These chunks are then transformed to the frequency domain and stacked as ``separate'' data streams within the MCMC sampler. Within each of the eight larger segments, we allow the merger time to vary only across that segment (i.e., the edges of the prior on the merger time are the edges of the 1/8-year time window). When it is time to evaluate the Likelihood for a sample in the MCMC run, we determine which 1/2-day window its merger time falls into. We then calculate the Likelihood using the associated 3/2-day window, centred on the 1/2 day in which the merger occurs. This, once again, allows us to focus on merger detections specifically. There are advantages and disadvantages to this which will be further discussed in Section~\ref{sec:disc}. 

This strategy will produce discontinuities in the Likelihood when the merger time crosses from one window to another. However, in practice these discontinuities create only small disturbances in the Likelihood. They are further minimized by including the 1/2 day before and after the window of interest. Therefore, this does not cause any noticeable issue in the sampling allowing this search process to proceed properly. 

Each of the eight larger pieces of data are analysed independently. The MCMC in each chunk is run until the Likelihood converges to a maximum value (remaining fixed for 500 sampler iterations) and every sample in the cold chain is within a few tens of that maximum value. The detected and optimal SNRs are then calculated. Right now, we apply a threshold of 20 to the smallest of these two SNRs. This choice is loosely based on what was found empirically in previous works~\cite[e.g.][]{Katz:2021uax}, but in the future we plan to further investigate this choice of threshold and its effect on the MBHB search results. Under an SNR of 20, we begin to confuse potential sources with a combination of the noise and the GB signals. However, as stated earlier, the loud MBHBs in the Sangria datasets allowed us to leave the resolution of this issue to later work. 

If the source found has detected and optimal SNRs greater than 20, the source is kept and added to a running list of found sources. These sources are then subtracted from subsequent runs. This process is repeated until each of the 8 larger-time pieces are unable to locate a source above the SNR threshold. 

This process usually results in finding more sources than are actually there. We prune our catalog for duplicate entries. The probability of an extreme match between two astrophysical MBHB signals is infinitesimally small; therefore, we assume that a high match between the signals indicates imperfect extraction of the first signal found, with a second or third signal completing the extraction process. The first source is usually very close to the true point with subsequent found sources attaching to a much smaller fraction of the signal. We keep, as is, only the highest SNR source within each of these multiple-source groups. After performing this pruning operation, we end up with the 15 (6) MBHBs that are truly injected in the LDC2A training (hidden) dataset, indicating its success. This will obviously need to be tested on more datasets in the future, but our initial tests give us confidence it will perform correctly. In addition to the sources of interest, this search also provides a set of samples that act as a starting point for each source in the larger global fitting run to come.

Following the culmination of the MBHB search pipeline, we run the global fit sampling operation without the GBs: the PSD, foreground, and MBHB parameters are allowed to vary until the PSD and foreground estimation properly accounts for the MBHBs removed from the residual. This process culminates when it converges to a maximum Likelihood value. Following this occurrence, GBs are added back in and the full global fit run commences. 

\subsection{MBHB Parameter Estimation}\label{sec:pe_mbhb}

The MBHB parameter estimation operation is more demanding than the PSD module, but not as difficult as the GB module. It runs in a very similar way to the PSD module: it uses ten temperatures spaced the same way as the PSD run; a similar update function to read in and out information; and a vertically arranged set of residuals and PSDs for marginalization, removing the permutation property of the ensemble tempering scheme (for computational efficiency). 

The main difference between the MBHB part of the sampler and the PSD module is that there are multiple MBHBs in each of the datasets. In this iteration of our global fit algorithm, we operate the MBHB portion as a fixed-dimensional sampler that assumes the search has found all sources of interest. In future work, this will be adapted into a trans-dimensional setup that is similar in nature to how we currently run the GB module.

We implement a special proposal move designed specifically for sampling across the MBHB sources. It is described in Algorithm~\ref{alg:mbh} in Section~\ref{sec:mbh_prop} in the Appendix. The main idea, like the global sampler as a whole, is to sample each source in a parallel tempered fashion while marginalizing over only the cold chain information from the other sources. 

We leverage the fast heterodyned Likelihood functions available in \texttt{bbhx}~\cite{bbhx_zenodo,Cornish:2021lje} for improved efficiency. For more information on heterodyning MBHB waveforms, see~\cite[e.g.][]{Cornish:2021lje, Katz:2021uax}. The MBHB proposal algorithm is given in the Appendix.

The ``inner proposal'' that operates each individual step of the sampling operation is chosen each step at random from the four proposals detailed in \cite{Katz:2021uax}. The main proposal chosen with $\sim90\%$ probability is just the regular stock stretch proposal from \texttt{eryn}~\cite{eryn_zenodo, Karnesis:2023ras,affine1}. The other three split the last 10\%. These are ``sky moves'' that leverage sky-mode symmetry by changing the sky-mode in which the point lies, while keeping all other parameters fixed. The three different moves include an only-latitudinal move, an only-longitudinal move, and a move that is both latitudinal and longitudinal. See \cite{Katz:2021uax} for more information on these moves. 

\subsection{Galactic binary search and parameter estimation}

By far the most complicated operation of our global fit pipeline is the detection and characterization of GBs. The complication is due to many aspects of the problem, including the large number of expected signals, a high degree of confusion between signals, and the fact that there is an unknown number of signals. 

To accomplish this task, following the lead of \cite{Littenberg:2023xpl}, we use Reversible Jump MCMC \cite{rjmcmc1} to characterize the number and type of signals detectable in the LISA data. In this section, we will describe the overall properties of the GB algorithm, detail the various proposals used in the GB sampler, and discuss how we proceed naturally through search and parameter estimation stages. 

With GBs in the global fit, there are three scales of concern. The smallest scale represents the individual sources (dimensionality, $D\sim8$). The largest scale involves all sources across the whole frequency band ($D\sim10^5$). The middle scale is the scale over which individual source posteriors interact with each other. While this varies across the LISA frequency band, this scale involves the interaction of $\sim2-50$ sources via the overlap in their frequency domain spectra ($D\sim16-400$).

The GB algorithm is designed to primarily handle the middle scale (source interaction) while successfully meeting both the small- and large-scale requirements. We break up the computation across the full band into a sequence of sub-bands whose widths are primarily determined by the widths of the GB frequency spectra. The sub-band to which a source belongs is determined based on its initial frequency parameter. With one year of observation, the sub-band widths employed 
are 128 frequency bins (each of width $1/T_\text{obs}$) at low frequencies ($<1$ mHz), 256 at middle frequencies ($<$10 mHz), and 1024 at high frequencies ($>$10 mHz). For more information on this construction, please see \cite{Cornish:2007if}. 

\begin{figure*}
    \includegraphics[width=\textwidth]{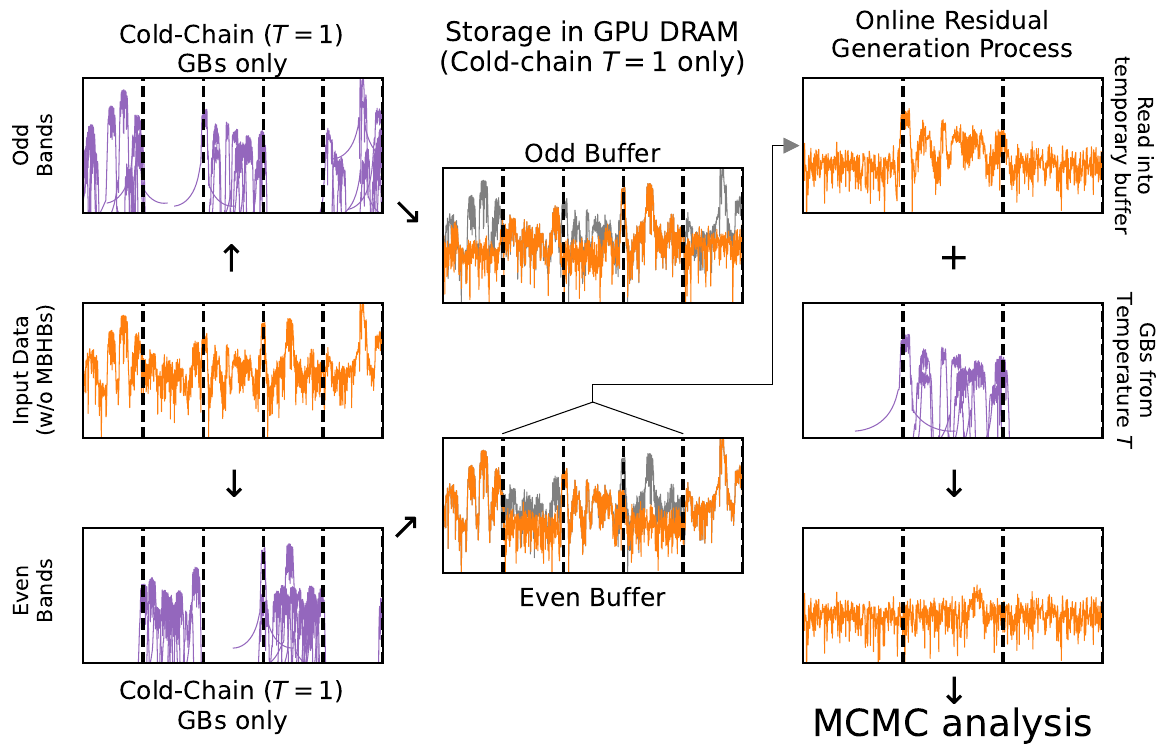}
    \caption{Illustration of the sub-band structure used to analyze the GB population, as well as the special storage of residual information for maximal memory efficiency. The left column, in the middle, shows the input data with the MBHB signals subtracted from it. The dashed black lines divide the window into sub-bands. These sub-bands represent the analysis unit of the GB algorithm. The sub-bands are run in parallel rotating between odd and even groups. During each specific sub-band analysis, we need information from three sub-bands: the sub-band of interest, the sub-band before, and the sub-band after the sub-band of interest. In the cold-chain ($T=1$), this is straightforward. At higher temperatures ($T>1$), the sub-bands before and after the band of interest need to come from the cold chain ($T=1$) because neighboring sub-bands will have generally different temperature ladders with only the cold-chain able to transfer information across the sub-band edge. If we stored three sub-bands for each sub-band of interest, it takes (per TDI channel) $\sim$3 $\times n_t\times n_w\times n_b\times$(length of each sub-band) $\approx$ 3 $\times n_t\times n_w\times$(data length) 16-byte numbers. This can quickly fill available GPU DRAM (device RAM). Instead, we store \textit{cold-chain} odd-band and even-band buffers that are read in at runtime and adjusted to the above three sub-band setup needed for analysis. This uses (per TDI channel) 2$\times n_w\times$(data length) 16-byte numbers, reducing the storage needed by a factor of $\sim n_t/2$. To form these odd and even buffers, we duplicate the input data (with MBHBs removed) and subtract waveforms belonging to odd \textit{or} even bands from the two instances of input data (purple waveforms in the top and bottom plots in the left column). The resulting buffers are illustrated above in the center column. The orange shows the stored residual and the grey shows the original input data to use as reference to see which sub-bands have been updated. This storage method separates into odd and even bands while maintaining any signal leakage across sub-band edges. The online residual formation process is shown in the right column. When running online computations on an odd (even) band at any temperature, the three sub-bands needed for the analysis are read-in from the even (odd) cold-chain buffer to a temporary GPU buffer. Then, the proper residual for analysis is formed by removing the waveforms belonging to the temperature, walker, and sub-band of interest.}
    \label{fig:band_diagram}
\end{figure*}

Within each sub-band is really where the MCMC-related operations are run. This means proposals, tempering, and accounting for source changes are all performed at the sub-band level. Within the ensemble setup, we leverage multiple walkers and multiple temperatures for each sub-band. However, there are a few important limitations. First, while sources are grouped into the sub-bands by their initial frequency parameter, their spectrum may extend beyond what we call the ``band edge'' of the sub-band. Figure~\ref{fig:band_diagram} shows the leakage of the GB template models into neighboring sub-bands: the purple template waveforms cross over band edges represented with black dashed lines. This is natural and represents our inability to specifically localize each source to its own set of frequencies. For this reason, sub-bands are run in an ``odds and evens'' way, where the odd-numbered bands are run first and even numbered bands are run second. This holds fixed sources with spectra that may leak across the band edge while sources within the band of interest are moved with MCMC operations.

We allow samples to go halfway into neighboring bands while the sampling within a band of interest is occurring. This prevents the effect of ``hard'' edges barring samples from moving across them in their initial frequency parameter. After a sequence of sampling operations within the band of interest, if a cold chain sample falls into a neighboring sub-band, the sub-band ownership over that sample is shifted to the neighboring band it now falls under. This treatment of these ``edge effects'' is similar to~\cite{Littenberg:2023xpl}.

Second, similar to the PSD and MBHBs, we want all residuals to be cold-chain information. In other words, we want all computations to be performed on residuals representative of the posterior distribution ($T=1$) which ensures the marginalization efficiency is maximized. For this reason, all sub-bands at higher temperatures are marginalized against neighboring sub-bands from their associated walker index ($w$) \textit{in the cold chain}. In the GB sampler, because the tempering is not as much of a computational bottleneck, we choose to do a full ensemble-permuted tempering setup. This helps to  filter the various sub-band states more efficiently to the cold chain while properly marginalizing over the various residuals and PSD information coming from the other parts of the global sampler. 

We run our sampler with 18 walkers in 24 temperatures, allowing for a large number of running sub-bands to better sift through all the signals and output the desired information. Our temperature ladder at the start ranges from 1 to infinity with the temperatures in the middle spaced geometrically by a factor of 1.2. These temperatures do adjust over time. Each sub-band is given an independent temperature ladder. This means that the sub-bands will generally have different temperature ladders that will ideally adjust to their differing characteristics. Only their cold chain information will be able to directly translate between sub-bands because the cold chain is the only temperature rung guaranteed to be at the same temperature as its neighbors at all times. 


As the sampling proceeds, a user-defined number of in-model proposals are performed consecutively (we use $\sim30$) followed by one reversible jump proposal and then temperature swapping. For in-model proposals, we use the group stretch proposal discussed in Section~\ref{sec:mcmc}.

For between-model moves, we use a mix of various proposals. The most basic RJ proposal is a draw from the prior distribution. We refer to this proposal as the ``Prior RJ'' proposal. It is the least informative proposal, but is helpful to ensure the entire space is explored. 

\subsection{Online updating of RJ sampling distributions}

In addition to the Prior RJ propsoal, we developed two proposals based on fitting Gaussian Mixture Models (GMM)~\cite{scikit-learn} to empirical sample distributions for resampling within the larger posterior estimation run. The first of the GMM-based proposals, the ``search RJ proposal'', searches every sub-band independently and simultaneously for the maximum Likelihood source that still remains to be subtracted from the best-fit residual. To do this, it runs a sampler that is fixed-dimensional, parallel tempered, and stretch-proposal-based. Once the highest Likelihood point still remaining in each sub-band is found, samples are generated around each point to build a distribution to sample from within the larger sampler. We fit each source separately to its own GMM and then combine them all into one larger GMM by assigning each source-specific GMM as having equal probability. More information about the algorithmic side of this process can be found in Section~\ref{sec:fitting} of the Appendix.

A similar process is performed for the other GMM-based proposal. However, rather than the inputs being points found from a search operation, the inputs are samples that repeatedly appear in the sampler. Once the sampler reaches a user-defined number of iterations, we go back roughly 10 (thinned) samples within each walker and perform a cataloging operation (see Section~\ref{sec:cataloging}). The samples that appear regularly are grouped into distinct distributions that are then fit individually with a GMM. Similar to the search RJ proposal, we combine all these individual distributions into one larger GMM by assigning equal probability to each individual GMM. We refer to this proposal as the ``refit RJ proposal.'' The main point of using this proposal is to ensure satisfactory mixing at higher temperatures by proposing points that have already been found, but may not appear in all higher temperature chains depending on their SNR.

\subsection{Transition from search to parameter estimation}\label{sec:search_to_pe}

At the start of the full global fit run, after the initial search for the MBHBs, overall instrumental noise and foreground confusion level, we begin the GB sampling run with special settings. The purpose of the initial search settings is to prevent a specific problem: when starting with zero GBs in the model, it is common for the sampler to fill itself with an excess of low-SNR sources while it takes time to find the louder sources. This is mainly caused by proposals like the prior proposal, since it uses an uninformed distribution. This can clog up the sampler with many unnecessary computations, significantly slowing its convergence. Additionally, even in a more refined state that is close to the ground truth, the sampler efficiency is highly dependent on the total number of templates currently being used. Therefore, many unnecessary and false detections cause the sampler to slow down. 

This is more of an issue when starting directly from one year of data. In \cite{Littenberg:2023xpl} the authors start with 1.5 months of data and gradually build up to use the whole year. They use the information from the previous duration to help the sampler more quickly converge for longer durations. This is a different but effective way to deal with the issue of small-SNR source buildup when beginning sampling runs at longer observation times. 

The approach we take to solve this issue, when using 1 year of data from the beginning, is to set up our RJ proposals in a specific way. We use only the search RJ proposal to both add and remove sources. This proposal is selected on average 80\% of the time. We do not use the refit RJ proposal until later in the search run. We use the prior proposal $\sim20\%$ of the time, but, importantly, \textit{we only allow it to remove sources}. Preventing the prior RJ proposal from adding sources prevents the addition of a large chunk of low-SNR sources. It also helps to ``prune'' the ongoing search by easily removing any sources that evolve through RJ or in-model moves to an SNR too low for them to remain in the sampler consistently. 

When the GB sampler begins, the search RJ proposal is finding the loudest sources in each sub-band. This distribution is then used to propose new points, which ensures the loudest sources enter the model first. As the search RJ proposal updates, it will use the best-fit data residual, which allows it to extract sources and move to lower and lower SNRs over the course of the run. To keep the large number of extra low-SNR sources out of the data residual, we use an optimal SNR limit of 7 in the search RJ proposal at the beginning.

The search operation finds $\sim600$ sources in its first iteration across all sub-bands. As the search proceeds, the number found in each iteration decreases. The end of the search phase comes when the maximum number of templates in the sampler converges. At this point, the settings are switched to those appropriate for full parameter estimation.

During parameter estimation an SNR limit of 5 is used for the search RJ proposal. The search proposal and refit RJ proposal are each used $\sim20\%$ of the time. The prior RJ proposal is used for $\sim60\%$ of proposals (both for addition and removal). These settings allow the sampler to focus on sources at the limit of detection, that may not have been found properly during the search operation, while still ensuring good mixing of chains at higher temperatures where the larger-SNR sources are added and removed.  

\subsection{Cataloging Operation}\label{sec:cataloging}

Our cataloging operation begins with an optimal SNR cut on all collected individual samples from the parameter estimation run: optimal SNR $>7$. We begin with one sample (meaning $\sim10^4$ templates from across the band) as the control group against which we make our initial overlap computations. We group all other templates in every other sample into a test group. For each binary in the test set, we find the maximum overlap with the sources of the control group. If this maximal overlap is greater than 0.9, the test set source is added to the group of the control source it matches with. We repeat this step with the next sample as the control and all other non-grouped sources as the test set. This continues until all samples have been treated as the control. Then, to remove any imperfections, we merge any groups that have an overlap greater than 0.9 with each other.

At the culmination of the cataloging operation, each of these groups is assigned a ``confidence'' value, $\mathcal{C}$. The confidence is the number of samples contained in a cataloged group divided by the total MCMC sample count. For louder sources with any appreciable SNR, this confidence fraction will be close to or equal to one because they will appear in every sample extracted from the global fit. We note the cataloging operation is not perfect. It requires making some set of assumptions and/or cuts. For this reason, a few samples may be missing from clearly detected sources, making their confidence values close to but not equal to one. 

For sources near the SNR detection limit $\sim7$, the confidence value may decrease because the RJMCMC algorithm may locate a model configuration that removes the marginally detectable source from the current sample. Confidence values for these sources can range from zero to one. Since full global fit runs and results have only recently been produced, the cataloging step--its choices and ramifications--are not well understood and are expected to improve in the future. Similarly, the relation of our confidence readings to these cataloging choices is a topic for future investigations. We also do not account for two or more cataloged binaries fitting one real source. Our initial checks indicate this does not happen often; a more detailed analysis of where and why it happens is left for future work. With these important points in mind, we believe our cataloging operation is good enough to prove our global fit algorithm's capabilities.


After the full operation is done and confidences are accounted for, the samples that make up the catalog are saved to HDF5 files to be read in and analyzed using \texttt{lisacattools}~\cite{lisacattools}. Many of the plots shown below were produced using this LISA catalog analysis package.

\subsection{Large-scale algorithmic setup}

The global fit is an extremely compute intensive procedure. For this reason, it is absolutely essential to have efficient sampling and Likelihood computations. We focus here on the large-scale setup. More specific information about the proposals, tempering, and the distribution fitting pipelines can be found in the Appendix.

Our overall goal was to use GPUs to handle as much of the global fit analysis as possible. The most important reasons for this are the efficiency of GPUs when it comes to handling highly-parallelizable problems and the fact that GPUs are expected to continue to improve at a faster rate than CPUs in the near future leading up to the LISA mission~\cite{gpu_better_efficiciency}. 

Every separate module (MBHBs, GBs, PSD, etc.) in our algorithm is run simultaneously using \texttt{mpi4py} \cite{mpi4py1, mpi4py2, mpi4py3, mpi4py4} for launching and communicating between all current computing processes. Each separate module uses a single GPU of its own (except for the MBHB search operation that uses as many GPUs as you give it). All of the main sampling modules run directly out of \texttt{eryn}~\cite{Karnesis:2023ras, eryn_zenodo} for sampling. The only customized parts of this are the special MBHB and GB proposals, as well as the GB tempering piece. They are, however, directly worked into the highly modular framework of \texttt{eryn} for adding customized proposals and tempering setups. 



Besides the specific proposals mentioned previously for the PSD, foreground, and MBHB analysis, their GPU-accelerated, vectorized Likelihood functions were important to the efficiency of this pipeline. The fast GPU-based PSD Likelihood was created specifically for this pipeline to ensure its computational and memory efficiency. As mentioned previously, the MBHB sampler uses Heterodyning~\cite{bbhx_zenodo,Cornish:2021lje} to improve its Likelihood evaluation time. 


For rapid evaluation of GB sampling operations while maintaining memory efficiency, we use a special setup for the storage of the GB residuals in the GB sampler. We store two sets of residuals: one for cold-chain residuals from odd numbered sub-bands and one set for cold-chain residuals from even-numbered sub-bands. However, both of these sets of residuals contain the original data with all MBHBs subtracted.  Therefore, to recover the full residuals across the entire band, it is necessary to add the two sets of residuals together and then subtract out the original data (with MBHBs subtracted). This process is illustrated in Figure~\ref{fig:band_diagram}.

The main idea here is that all sub-bands within the analysis need information only from the neighboring \textit{cold-chain} sub-bands. To illustrate this usage, imagine we want to operate at temperature $t$ for walker index $w$ in sub-band index $b$ which we take to be odd. We read in sub-bands $b-1$, $b$, and $b+1$ from the even-numbered cold-chain residual of walker $w$ into a GPU buffer. The only template waveforms subtracted from band $b$ (which is odd) in the even-numbered cold-chain residual are sources in the neighboring bands that leak over into the band of interest. We subtract from the buffer residual the waveforms contained in (walker, temperature, sub-band)=$(w, t, b)$. This produces, in the buffer, the full residual at the start of sampling with all current sources in the sampler removed from the sub-band of interest. This process is illustrated in Figure~\ref{fig:band_diagram}. This simulates the sub-band of interest at each temperature by reading in the proper cold-chain residuals of its neighbors and filling information in the sub-band of interest with its own sources at its specific temperature. This means we do not need to store any of the higher-temperature residuals in GPU global memory. Memory optimization on GPUs is very important. If we stored every higher temperature residual, the GPU memory burden would be $n_w\times n_t\times\text{(2 channels)}\times\text{(memory size of each residual)}$. We reduce this by a factor of $n_t / 2$ since we have 2x the cold-chain residuals. 


The alternative easier choice is to store 3 sub-bands worth of information ($b-1, b, b+1$) for every $(w, t, b)$ combination (with $b-1$ and $b+1$ coming from the cold chain). This would cost $n_w\times n_t\times n_b \times \text{(2 channels)}\times\text{(memory size of 3 sub-bands)}$, which is three times the cost of storing all higher-temperature residuals. 

When the proposal operation (see Section~\ref{sec:gb_proposals} in the Appendix) on the GPU is complete, the odd-numbered residual will be updated with the new sources in (walker, temperature, sub-band)=$(w, 0, b)$.

\section{\label{sec:res}LDC2A Results}

Our analysis of the LDC2A data is presented below. Overall, our pipeline performed well. It found, extracted, and characterized resolved signals of the expected types; characterized the detector noise and confusion noise foreground; and accomplished all of this 
without any human intervention. We ran our pipeline for $\sim7$ days on 4 NVIDIA A100 GPUs and $\sim12$ CPUs. 

\begin{figure*}
    \includegraphics[scale=0.85]{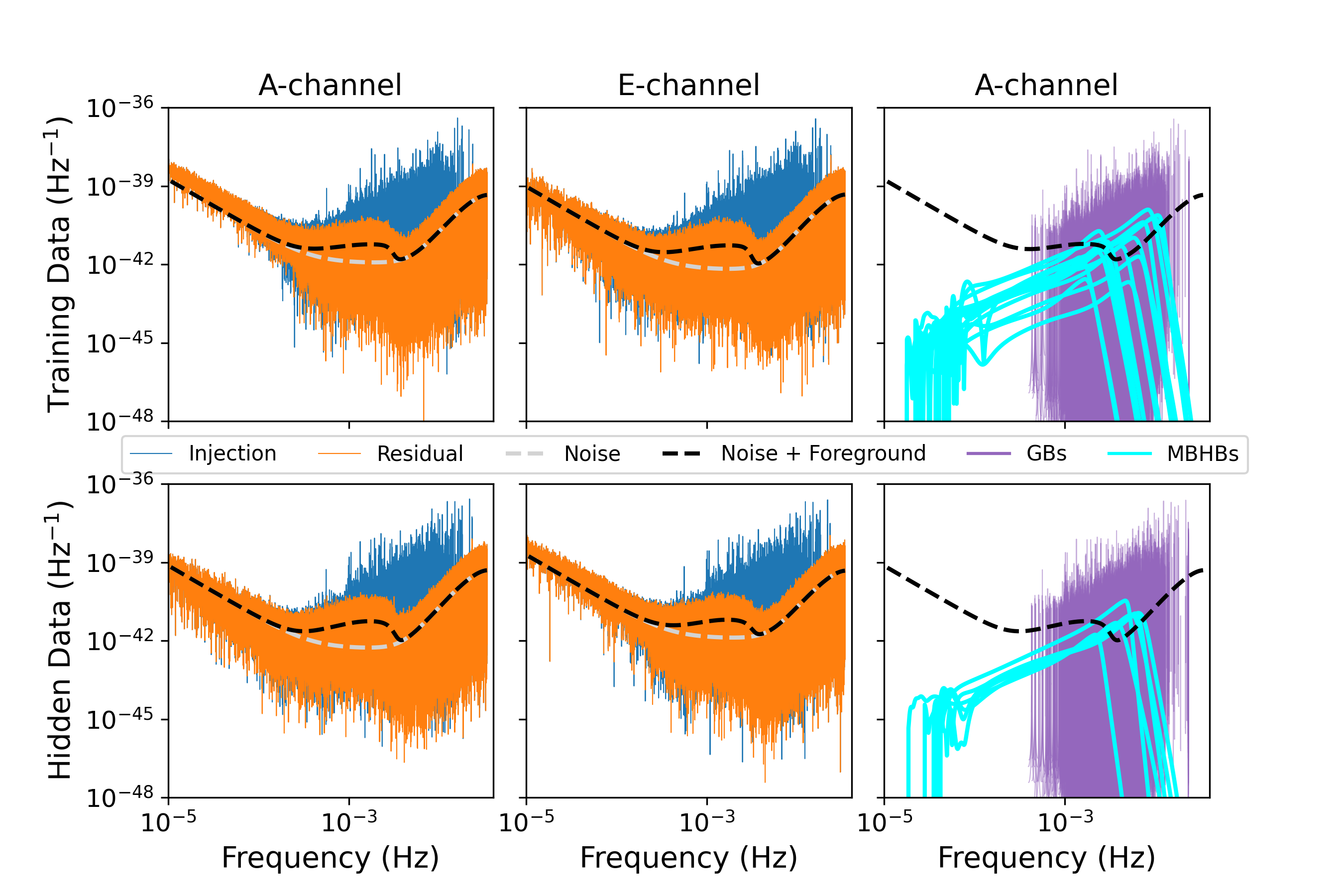}
    \caption{Comparison of injected (blue) versus residual (orange) data streams. The best fit noise profile is also shown with a black dashed line. The gray dashed line represents only the instrumental noise contribution. The difference between the black and gray dashed lines is the galactic confusion noise. The left and center columns are for the A and E channels, respectively. The column to the right shows all template waveforms subtracted to make the given residual, with GBs shown in purple and MBHBs shown in cyan. The upper row is the training dataset. The bottom row is the hidden dataset.}
    \label{fig:residual}
\end{figure*}

Figure~\ref{fig:residual} shows the high level picture of the source extraction and noise analysis. It shows the residual compared to the injection in the A channel (left) and E channel (center), as well as the template waveforms for the resolved sources subtracted to form those residuals (right). The training and hidden datasets are shown in the top and bottom rows, respectively.

The residual plot shows that we were able to extract all appreciable signals (note, the line in the residual spectrum at the right of the plot is an artifact of data generation and not a missed GB signal). This is true for both the training and hidden datasets. We also found and extracted all 15 MBHB signals from the training dataset, with no false or missed detections. For the hidden dataset, we extracted all 6 MBHB signals with no false or missed detections. 

The low-frequency noise fit for the A-channel (E-channel) in the training (hidden) data shows a small inconsistency with the residual data. We believe this is due to two factors. First, the injected noise model contains acceleration noise, Optical Metrology System (OMS) noise, and backlink noise (see~\cite{Bayle:2023qfo} for more information on these noises). Our fitted model given in Equation~\ref{eq:psd} only contains two of those noises: OMS noise and acceleration noise. This causes a mismatch in the noise model at low frequencies. This was due to a previous oversight related to the LDC documentation~\cite{ldc}. Second, our pre-processing operation did not perform the necessary detrending of the time-domain data prior to transforming the data. We have verified that detrending by removing a linear fit to the data over time removes much of this issue. This does lead to a small increase ($\sim100-300$) in the number of detected binaries and will slightly affect the individual posteriors of various sources. However, we did not rerun our entire pipeline because the observed changes were small enough that the necessary computational expense was not worthwhile.

\subsection{\label{sec:mbh_res}Massive Black Hole Binary Posteriors}

\begin{figure*}
    \includegraphics[scale=0.7]{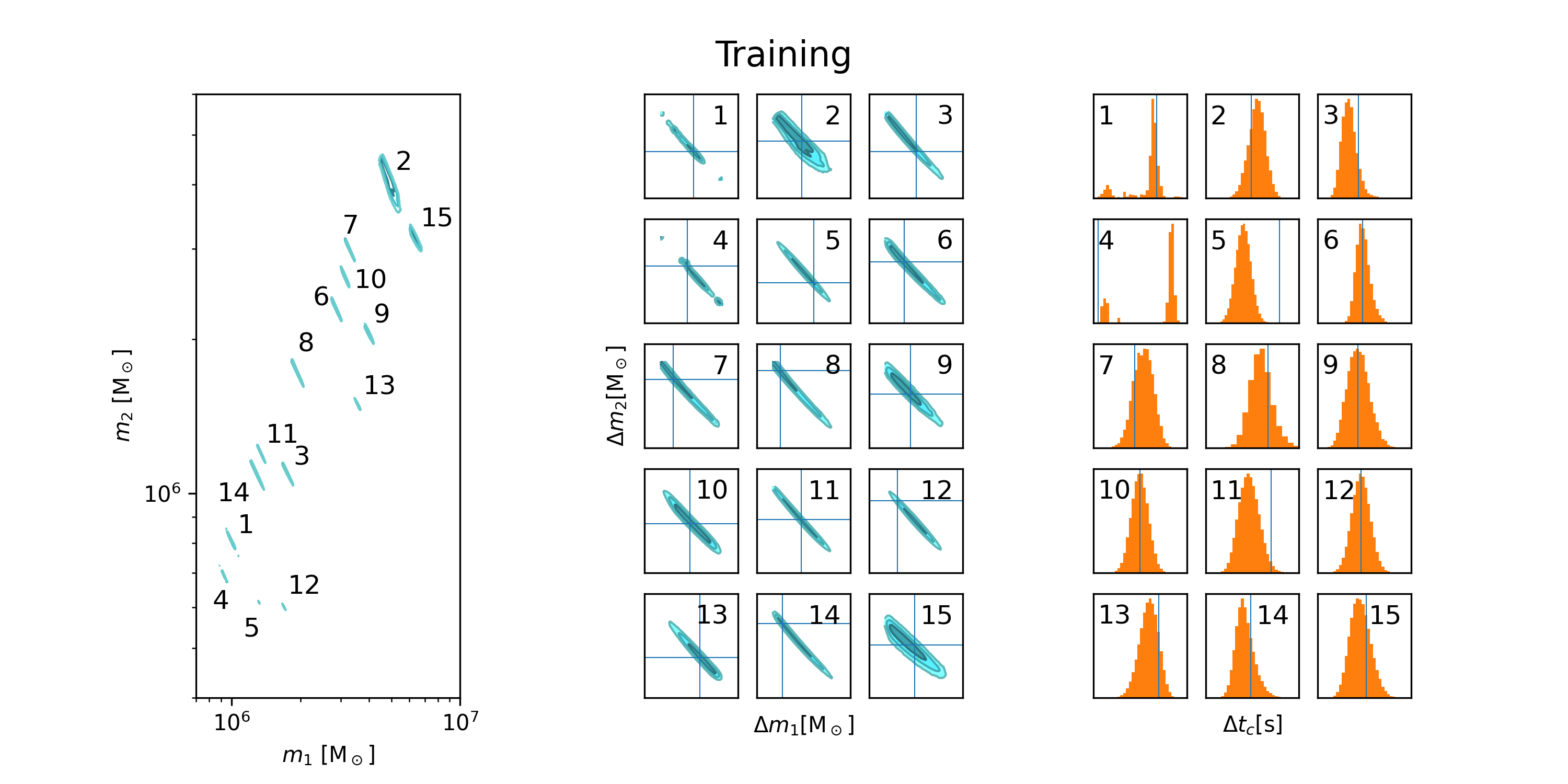}
    \includegraphics[scale=0.7]{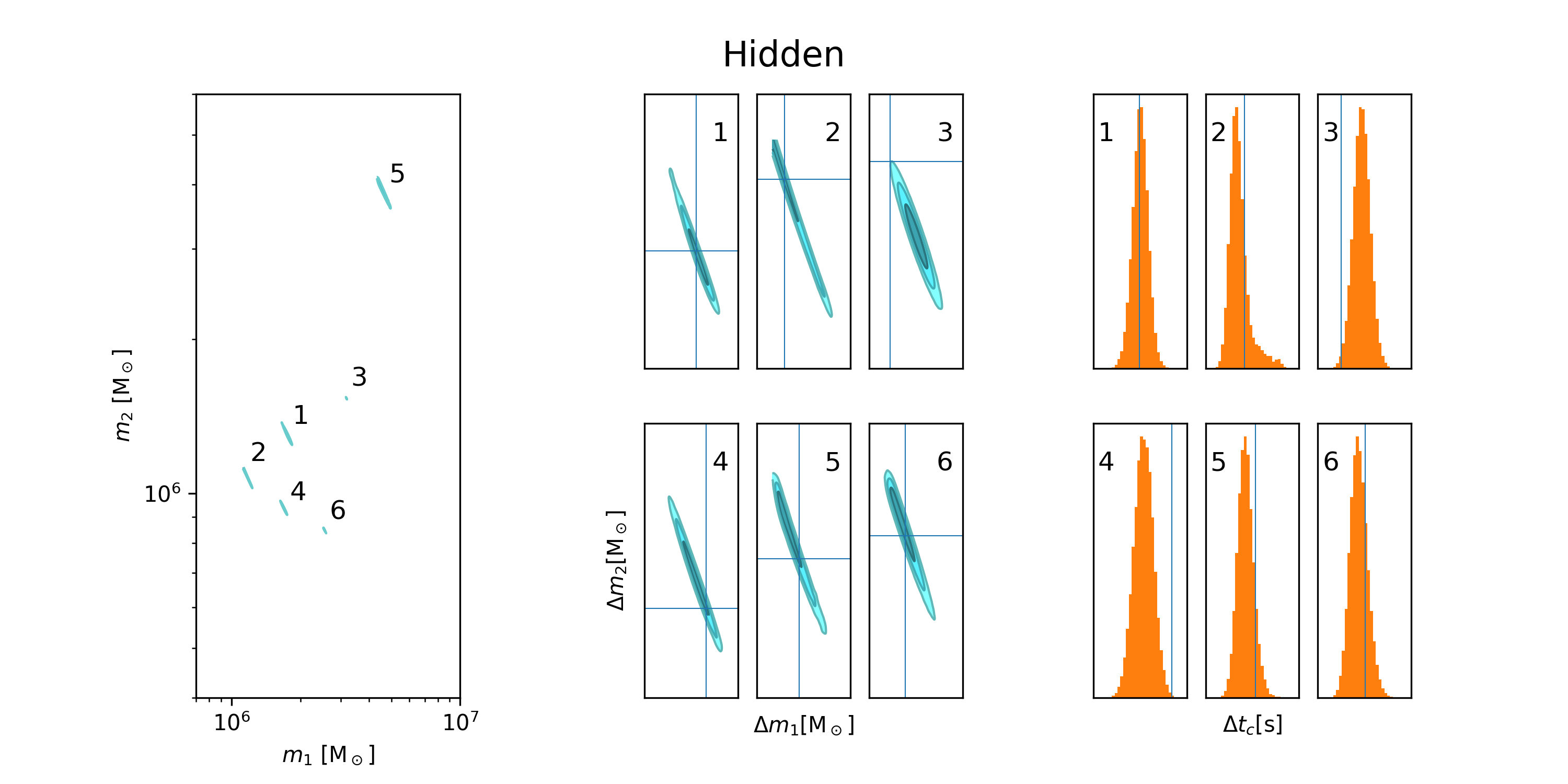}
    \caption{Posterior distributions for the MBHB sources. The left hand column shows posteriors on the masses of the binary components, including contours at $1$, $2$, and $3\sigma$ credibility. Sources are numbered in order of their merger time. The middle column shows the same results, but now zoomed-in and expressed as a difference to the injected value. The injected values are marked by blue lines. The right hand column shows 1D marginalized posterior distributions for the coalescence time. Injected values are again marked with blue lines lines. 
    The top/bottom row of plots shows results for the training/hidden dataset.}
    \label{fig:mbh_post}
\end{figure*}

The posteriors on the MBHBs are shown in Figure~\ref{fig:mbh_post} for both the training and hidden datasets. In this figure, we focus on the observed masses and detector-frame merger time information to concisely represent our results. The single-source posterior windows are labeled in order of coalescence time. Full posteriors can be created and analyzed for all sources by anyone using the \texttt{lisacattools} software package~\cite{lisacattools} and our final catalogs~\cite{katz_training_data, katz_hidden_data}.

Both black hole masses lie within the $2\sigma$ credible interval for all injected sources, with the exception of 2 sources (one in each of the training and hidden datasets), for which the true masses were within $3\sigma$ credibility. 
Most source coalescence times show unimodal Gaussian distributions with widths of the order of a few tens of seconds with the injected value lying within the posterior bounds. The fifth source in the training dataset contains the true coalescence time far into the tail of its distribution. The first and fourth sources in the training dataset show multimodal posteriors. This is likely caused by the same effect in both cases, but for the first source the larger of the posterior modes contains the true time of coalescence while for the fourth source, it is the smaller of the posterior modes. 
Similar results were obtained by other groups analysing this dataset~\cite[e.g.][]{Littenberg:2023xpl}. It is currently uncertain what causes these additional modes (separated by $\sim{10}$s) and will be a topic for further analysis in the future.

\subsection{Recovery of Galactic Binaries}

Our algorithm collected samples for and cataloged $\sim12000$ Galactic Binary sources including known (or verification), detached, and interacting binaries. Approximately 8000 of these were detected with high confidence ($\mathcal{C}>0.9$). All of these sources are included in the same Galaxy sampling module. Known GBs are found and characterised in the same way as unknown GBs. We do not specifically search for them, but we do use them as markers of the performance of the algorithm. 

\begin{figure*}
    \includegraphics[scale=0.8]{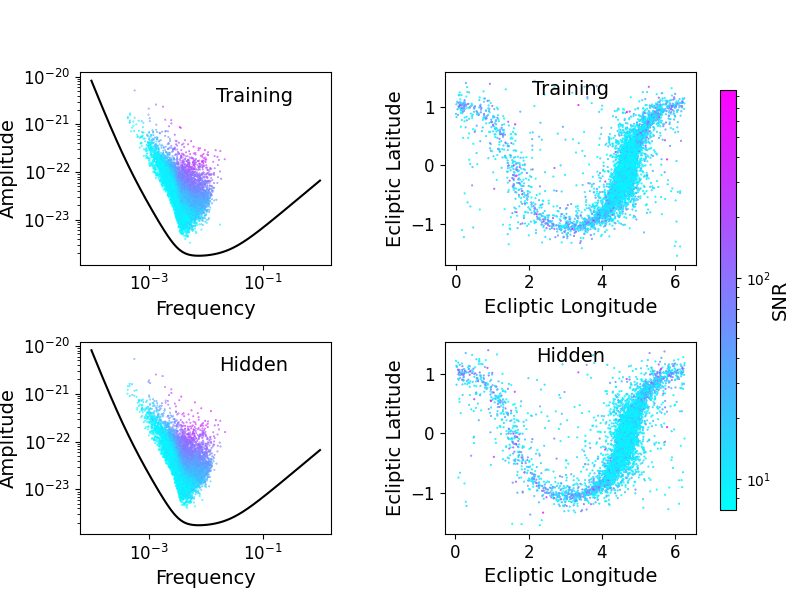}
    \caption{Point estimates for the cataloged population of GB signals observed in our global fit. The left and right plots show point estimates colored according to their SNR for amplitude versus frequency and ecliptic latitude versus ecliptic longitude, respectively. These point estimates are the median frequency sample for each cataloged source. The training (hidden) dataset is given in the top (bottom) row.}
    \label{fig:gb_det}
\end{figure*}

Figure~\ref{fig:gb_det} shows information about the detected population of GBs from the training dataset after cataloging. The $\mathcal{A}-f$ plot is color-coded according to the SNR of the detection. This plot also shows the expected dearth of sources between the sensitivity curve and the detected population. This difference is the characteristic confusion noise. The sky location point estimates are also given in the same figure. The sources congregate as expected along the Galactic plane. 


\begin{figure*}
    \includegraphics[scale=0.6]{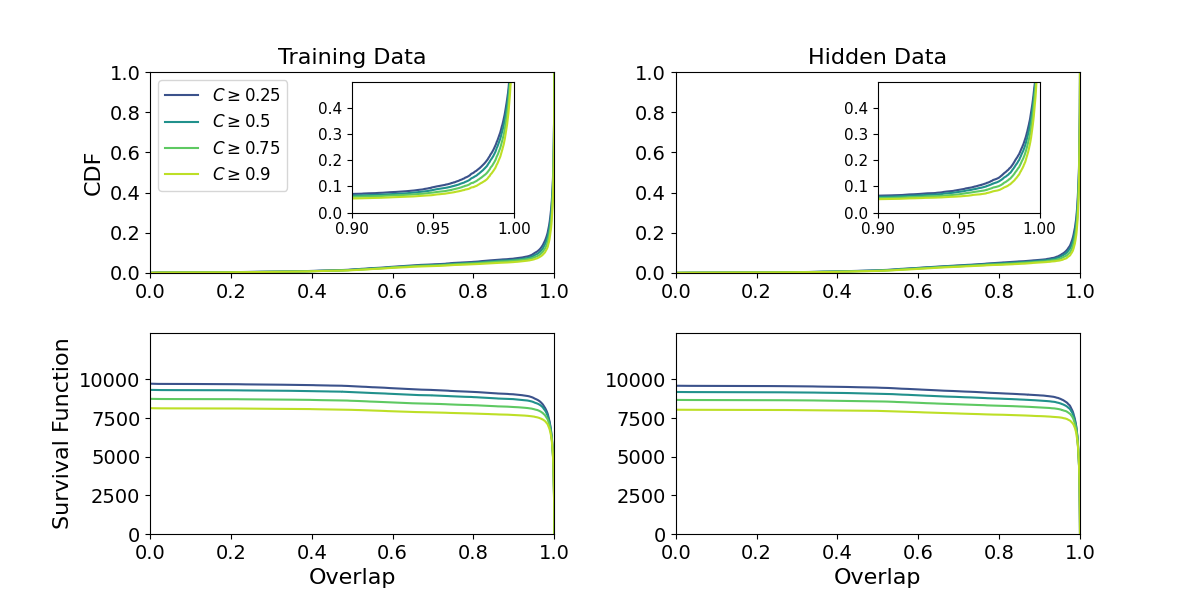}
    \caption{The CDF and survival function for the best match of a source within the GB output catalogs against the injected population. The injected population consists of all sources with optimal SNR $>7$. The lines are separated by their minimum confidence value, $\mathcal{C}$. The confidence is the number of samples in a catalog group divided by the total number of samples. As the confidence value decreases, more sources are included. The training dataset is shown on the left and the hidden dataset is shown on the right.}
    \label{fig:cdf_comp}
\end{figure*}

Our comparison to the injection catalog for each dataset is shown in Figure~\ref{fig:cdf_comp}. This plot is similar to Figure 13 in~\cite{Littenberg:2023xpl} and the right side of Figure 5 in~\cite{Strub:2023zxl}. Since the injected catalog technically has $\sim10^7$ binaries, an \textit{optimal} SNR cut of 7 is applied. This narrows the tested injection set to $\sim10^4$ injection sources to compare against our cataloged source samples. We take every sample contained for all sources in our catalog and compare them all against each source in the injection set. The maximum overlap achieved per injection source is plotted on the horizontal axis.


The cumulative distribution function (CDF) of this maximal overlap against each injection source is shown in the top plots (training on the left and hidden on the right). Their associated survival function is shown in the bottom plots. The curves are also labeled by their confidence, $\mathcal{C}$, i.e., the fraction of MCMC samples that contained a given source (in the range [0, 1]). As a reminder, the choice of a detection statistic in RJMCMC is nuanced and depends a lot on choices made during cataloging, especially on the margins. Understanding and optimising this is a topic for future discussion. 

Our algorithm performed similarly for hidden and training sets with a $>90\%$ match rate for both runs. The confidence lines are quite close together indicating that most of the samples have confidences greater than 0.9. 

\begin{figure*}
    \includegraphics[scale=0.65]{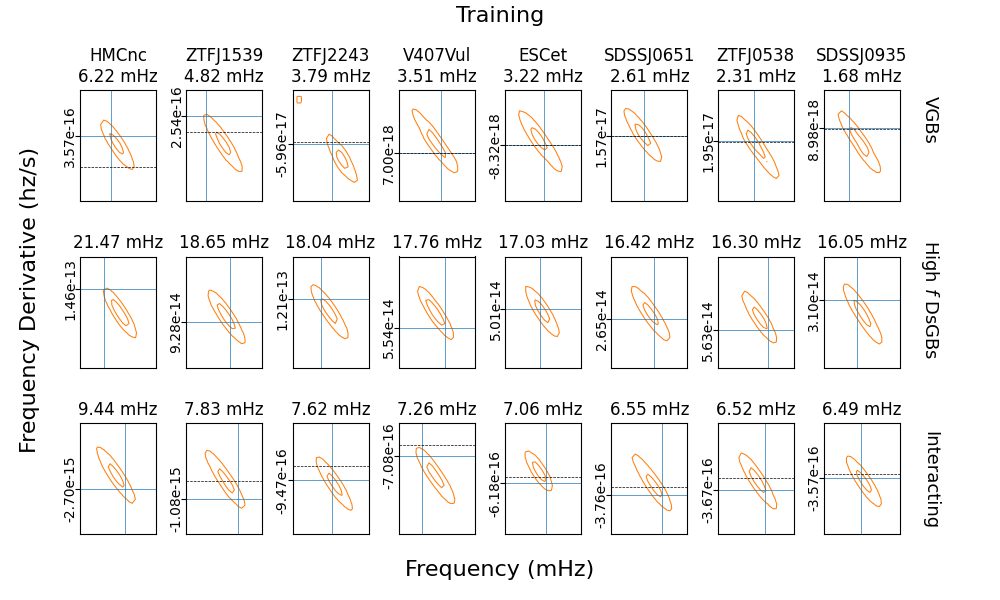}
    \includegraphics[scale=0.65]{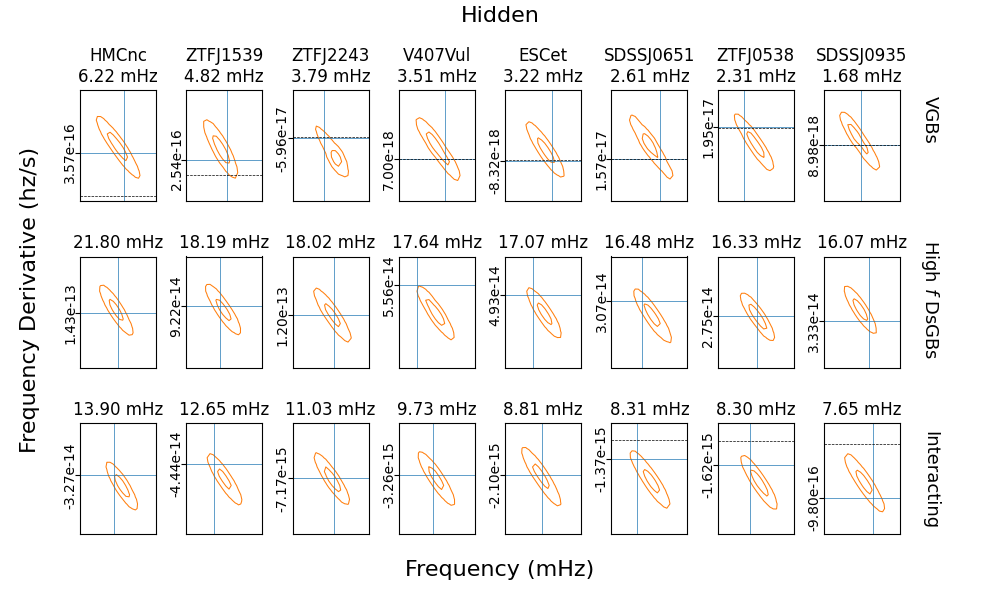}
    \caption{Marginalized 2D posterior distributions in the $f-\dot{f}$ plane. The top row represents the highest frequency VGBs arranged in order of decreasing frequency. These VGBs interact with other surrounding sources. The VGB parameters are not inferred using a targeted search as in \cite{Littenberg:2023xpl}, but simultaneously fitted with the remainder of the population. The middle row shows posteriors for the highest-frequency detached binaries in the randomly generated galactic injection population. These detached sources all occur at frequencies high-enough to avoid confusion with other detectable sources. 
    The bottom row shows posteriors for the highest frequency interacting GBs with $\dot{f}\neq\dot{f}_\text{GW}$. The orange contours show the $1\sigma$ and $2\sigma$ contours with the blue solid vertical and horizontal lines representing the injection values from the input population. The injected $\dot{f}$ value is given in units of Hz/s along the vertical axis. The horizontal dashed black line is $\dot{f}=0$, which gives a sense of the scale and constraint on $\dot{f}$. The training dataset is the top grid of plots and the hidden dataset is shown in the bottom grid of plots.}
    \label{fig:f_fdot}
\end{figure*}

With greater than 8000 detected sources of high confidence, it is difficult to analyze a large number on an individual basis. Therefore, we try to pick sets of sources that can help to indicate how well the sampler is performing. The three groups we have focused on are high-frequency detached GBs, high-frequency interacting GBs, and the known (or verification) GBs. Groups of eight $f-\dot{f}$ 2D posteriors for each of these three categories are shown in Figure~\ref{fig:f_fdot} (for both datasets).


The first group in the top row are the known VGBs. We have chosen the eight highest-frequency VGBs with $SNR \gtrsim15$ arranged in descending order according to their frequency. These sources all exist in regions with some level of confusion. HMCnc is above the main confusion noise frequencies, yet it still suffers from confusion with nearby detectable sources. The rest of the VGBs are all at frequencies where the confusion noise is expected to play an important role.

In~\cite{Littenberg:2023xpl}, they perform a targeted search for these sources by including direct information from EM observations. While this is a good idea in reality and will most likely be the method used during the final LISA data analysis, we chose to analyze these sources as a part of the larger population without any targeted assumptions. This allows us to test the performance of our algorithm against a specific subset of known sources and compare our posterior distributions to \cite{Littenberg:2023xpl} to understand how deliberately targeting the VGBs affects their constraints.

In Figure~\ref{fig:f_fdot} we  include a horizontal dashed line to indicate $\dot{f}=0$. For HMCnc and ZTFJ1539, the injected value is well separated from $\dot{f}=0$ and the posterior excludes this value. This indicates a confident detection, which will provide a stronger constraint on the source chirp mass. As we move to lower frequencies, shown on the right of the figure, the $\dot{f}=0$ line and the injected value converge until $\dot{f}=0$ is well contained within the posterior uncertainty. This is an indication that there are weak constraints on $\dot{f}$. 

The middle row of Figure~\ref{fig:f_fdot} shows posteriors for the eight highest-frequency detached GBs. As a reminder, ``detached'' means that the only driving force of the gravitational-wave chirp is General Relativity. These sources are very useful to understand how well the sampler's within-dimensional or in-model moves are working. These sources are found at frequencies where the density of sources is low enough that we do not expect them to overlap. Therefore, we can check our global fit posteriors against single-source, fixed-dimensional MCMC runs. We have verified that this procedure produces identical posteriors.

GBs at high-frequency will also have strong constraints on their $\dot{f}$ parameter. For this reason, none of the plots in the middle row of the Figure show a line at $\dot{f}=0$ because it is beyond the range of the axes. 

The final row shows posteriors for the interacting GBs. While these sources have a strong astrophysical basis~\cite[e.g.][]{Nissanke:2012eh}, the actual $\dot{f}$ values injected into this population were assigned ad hoc rather than with a specific parameterized model, largely due to the lack of a robust astrophysical prediction for this population.. 
Nonetheless, we were able to constrain $\dot{f}$ values well enough to show at least a few of these sources have negative chirps. 

As we move towards the right of the bottom row in Figure~\ref{fig:f_fdot}, to sources at lower frequency, the constraints weaken, but still show potential detections of a negative chirp, and hence a difference to the predictions of vacuum GR. We are unable to translate this into a meaningful bound on the astrophysical parameters or chirp mass because of the ad hoc nature of the injected population. However, this should be pursued with future datasets. 

\subsection{Recovery of instrumental noise and confusion foreground}


The fitting of the instrumental and confusion foreground noises is unique within the global fit in that the algorithm is fast and efficient, but it is slow to converge because it is highly dependent on the extraction of all detectable GBs. In other words, the noise PSD and confusion foreground information only converge once the galaxy fit has converged. As the GBs are found and removed from the residual, the foreground noise estimate decreases gradually down towards its final state.

As discussed at the beginning of this section, there was a mismatch in the low-frequency noise fitting due to incorrect detrending of the data and using a fitted noise model that did not match the injected model (missing the backlink noise component). This most likely affected the foreground estimation as well, but it is hard to determine by how much. We leave this to future work.

The posteriors on all of the instrumental noise and foreground signal parameters were found to be fairly Gaussian in both runs. The uncertainty in the total PSD was very small, such that when plotting the PSD as a function of frequency, as in Figure~\ref{fig:residual}, the confidence intervals are smaller than the width of the plotted line. However, individual parameters in the PSD model are less well constrained. We were able to constrain the confusion foreground amplitude to $\sim5-10\%$. We believe this is larger than expected \cite[e.g.][]{Muratore:2023gxh} due to parameter correlations built into our model parameterization clearly seen in the foreground posterior distribution. Samples for the instrumental noise and foreground parameters over the course of the search and parameter estimation runs are included in our output catalogs~\cite{katz_training_data, katz_hidden_data}.

\subsection{Algorithmic performance}\label{sec:alg_perf}

In this section, we will discuss some initial findings about the algorithmic performance of each module of the pipeline. The first three segments of the pipeline -- initial instrumental and confusion noise fit, MBHB search, and MBHB plus PSD information mixing -- all occur fairly rapidly. 

The initial overall noise PSD fitting takes under 30 minutes. There are some potential ways to improve this timing, but given it is a small fraction of the whole pipeline, only a small amount of initial effort was put into optimizing this segment. The MBHB search, which results in a reasonable posterior estimate of each source, takes $\sim$hours with roughly eight processes running simultaneously over four GPUs. This could also be improved in various ways, but 
it is still a small percentage of the overall pipeline run time. The mixing of the MBHBs and PSD fits also takes less than one hour. 

The Galaxy analysis is the clear driver of the global fit computational cost. The PSD information and MBHBs need time to converge to a posterior, but in the absence of GBs, this would take $\lesssim1$ day. 
The $\sim7$ days of total runtime is to allow for convergence of the Galaxy analysis. The most challenging part about the GB analysis is the scale: with $\sim10^4$ sources, the overall runtime can be slow, even when the evaluation cost per source is extremely efficient. 
Our approximate MCMC generation efficiency is 1 \textit{proposal} per binary source per $\sim1 \mu$s, with actual waveform generation occurring at a rate of 1 waveform computation per $\sim0.3\mu$s. These timing estimates occur when performing a large number of these computations in parallel as we do in our algorithm.


In different frequency regimes, the algorithm faces different difficulties. At low ($\lesssim1$mHz) or high ($\gtrsim10$mHz) frequencies, there are fewer resolvable sources which means these regimes generally do not greatly contribute to the overall runtime of the algorithm. At low frequencies, the primary complication is the confusion foreground. At high frequencies, the sources are well separated, which is generally easier to deal with than the confused regimes. 

The middle frequency band (1mHz $\lesssim f \lesssim10$mHz)is the most computationally challenging. The foreground confusion noise is above the instrument PSD at the start of this middle frequency window. It then falls below the PSD around $\sim4$mHz. While the confusion foreground decreases over this range, the confusion between \textit{detectable} sources increases, reaching its peak around $\sim3-5$mHz and then decreasing again before sources become more isolated at $\gtrsim10$mHz.

\section{\label{sec:disc}Discussion}

The LISA global fit is an extremely complex task that is difficult to accomplish. Here we have presented a new global fit pipeline capable of extracting multiple sources of different types, and a fit to a stationary noise profile. Our pipeline is not the only global fit algorithm that has been developed, but 
it is extremely useful to now have multiple global fit algorithms within the community. Having multiple algorithms will help with research into many open questions pertinent to the planning of the LISA mission. One example is how all the output catalogs produced by the multiple pipelines that will be run as part of the ground segment will be conglomerated into a single output catalog to be released by ESA. This will require making a variety of mixing assumptions and choices between the outputs of the individual global fit algorithms. 

As well as a number of minor differences, our algorithm adds three main novelties in comparison to~\cite{Littenberg:2023xpl}: i) GPU acceleration, ii) ensemble sampling, and iii) the online fitting of RJMCMC proposal distributions.

\subsection{GPU Acceleration}

The GPU set up has improved computational efficiency relative to~\cite{Littenberg:2023xpl}. It is difficult to directly compare the timing between the two implementations. Our implementation starts at one whole year of data analyzing it all at once, while~\cite{Littenberg:2023xpl} builds up the analysis over time, using information from early times for later analyses. This time build up is a more natural way to design a global fit algorithm given that the mission will be producing data for several years, and we do not want to wait until the end of data taking to start an analysis. In the future, we will adapt our algorithm to build up over time as well. It is hard to predict how this will affect the timing of the algorithm, so that comparison is left for future work. 

If we compare with what we have, $\mathcal{O}(7)$ days on 4 GPUs, to $\mathcal{O}(5)$ days on $\mathcal{O}(1000)$ CPUs, our algorithm is about $\sim5\times$ more efficient in terms of power usage and cost (multiplying power usage by a flat price per kWh). This is encouraging and we hope it will continue to improve in the future. 

When building algorithms for GPUs or CPUs, there is a trade off between performance, availability, and ease of use. The goal of this code is to have building blocks that work for both machine types (most likely optimized for the GPU), with wrappers that run them in the desired fashion. 
Being CPU/GPU agnostic will also aid in development. 

The GPU development is necessary in our opinion because it is clear that algorithms designed for the GPU will be more energy efficient. This is already true at the present time and, as we move into the future, it is expected that GPUs will advance at a much more rapid pace than CPUs, meaning the efficiency difference between the algorithms will most likely increase further. Our Galaxy sampler generally saturates the GPU with computations. Therefore, newer generations of GPUs like the NVIDIA H100 should improve the evaluation time of our algorithm because of the greater number of CUDA cores available. 

Currently, computing facilities with CPU hardware are more widely available than facilities with GPUs. However, the GPU demands of our algorithm are not great, and most new computing clusters that have $\mathcal{O}(1000)$ CPUs also have multiple modern GPUs available. 


\subsection{Ensemble sampling}

Ensemble sampling is a very important aspect of our algorithm. It allows for two main advantages: multiple cold-chains generating samples in a more parallelizable fashion and better mixing between the temperatures due to the random permutation of the walkers when swapping between temperatures. The full ensemble, the total number of walkers in all temperatures, provides a large number of computations for building up sub-band searches and proposals that feed down to the cold-chains. 

It is hard to assess or quantify the impact of ensemble sampling and tempering on the efficiency of this sampler. Qualitatively, it seems it can only help. At a minimum, more parallelization is possible since all walkers in all temperatures have Likelihood computations that are all independent of one another. The improved parallelization will also strongly benefit as GPU technology and capacity improves.

In the MBHB and PSD sampling modules, we currently do not employ ensemble tempering. This is because it requires the computation of extra Likelihoods that we have currently deemed to be unnecessary. However, in the future this may be useful to improve convergence. We could perform an ensemble tempering move periodically after a specified number of iterations, while maintaining the ``vertical'' tempering structure during the majority of proposals. This would surely help the mixing. Future work will determine its actual impact on efficiency and sampling quality, and optimize the frequency of vertical and ensemble tempering proposals. 

\subsection{Single source search-to-parameter estimation information transfer}

Another important piece of our algorithm is how we build up the galaxy population. We start with a single-source MCMC search that gives a posterior estimate. This estimate is then transformed into a proposal with a GMM and sampled directly as an RJMCMC proposal distribution. 

This search-to-proposal pipeline was essential to the success of our Galaxy search due to the fact that we analyze the entire year of data at once. If we relied on basic RJMCMC proposals, it would take prohibitively long to converge because all of the loud sources would be fit with multiple templates at the beginning. In \cite{Littenberg:2023xpl}, they avoid this issue by building up over time. This time build-up method finds louder sources first when they are the only detectable sources and then adds quieter sources as they become detectable over time. 

In addition to the careful addition of sources to build up the galaxy solution, this search-to-proposal method ensures that we will never leave a source of appreciable SNR in the data. It will always locate the best fit template that remains in the data residual (if there are any left). 

This method can be extended beyond the galaxy fit. It is at its core a general method for adding sources into a larger global fit running on an RJMCMC backend. This would allow for much greater flexibility compared to fixed source searches like the MBHB search we currently use. It would be more capable of ensuring sources are removed down to the lowest detectable SNRs that are only found as the full pipeline converges. 

Other types of astrophysical sources, such as EMRIs, stellar-origin black hole binaries, and MBHBs, could also be treated using this framework. We can sample them with a single-source search and parameter estimation algorithm to get an initial estimate for the source posterior. We will then fit these posterior estimates with GMMs or normalizing flows~\cite[e.g.][]{Korsakova:2024sut}. Once they are transferred into the global fit and allowed to fit simultaneously with all other sources, the initial estimated posteriors will be refined. The usefulness of this method cannot be overstated as we add more sources and source types in the future. We expect it to play a key role in ensuring our global fit has the full range of required capabilities.


\subsection{MBHB Recovery}

We recovered all 15 (6) of the MBHB sources in the LDC2A training (hidden) dataset. This search was done in a rather ad hoc manner and could be improved. The search currently uses an SNR cut of 20. However, this SNR is computed against a PSD that is far too high since this computation comes before fitting any GBs. Therefore, this SNR 20 cut corresponds to a much higher optimal SNR. 


Having this higher-than-expected effective PSD at the beginning of the MBHB search helps to eliminate spurious sources at the cost of sacrificing our sensitivity to low-SNR MBHBs. In the future, we will have to improve this. One potential solution is to use the same system of adding GBs for MBHBs (as mentioned in the previous section). This will allow for the search and addition of MBHBs with RJMCMC rather than the fixed search we currently employ. This would mean that as the overall noise PSD converges and GBs are subtracted, we will have the ability to add low-SNR MBHBs that cannot be found without near complete GB subtraction. 

In our current implementation that uses the entire year of data in one analysis, we split up the MBHB search computation into separate frequency windows and analyze them simultaneously. Within this setup, we have not considered the problem of pre-merger detection of MBHBs. The inclusion 
of this feature of the global fitting process is left for future work. 

In the future, there are also obvious improvements to be made in terms of the quality and type of MBHB waveforms used in the fit. The LDC2A has just the $l=m=2$ mode of the PhenomD waveform~\cite{Khan2016, Husa2016}. In the near future, there are plans to re-release the LDC2A dataset with higher-order modes included using the PhenomHM waveform~\cite{London2018}. While still not the most up-to-date or complete waveform available (it lacks features such as precession and eccentricity), it will be slightly close to what will be required for the final 
global fit algorithm. Although these waveforms are more expensive to generate, higher modes are expected to make the search and parameter estimation of these important sources easier while producing posteriors that are much more constrained in most cases~\cite[e.g.][]{Katz:2021uax}. Therefore, we expect the algorithm will perform as well or better when this waveform improvement is included. The inclusion of precession is also not expected to change this conclusion much, other than adding further cost to the waveform evaluation. Adding other physical effects like eccentricity are likely to have a bigger impact, but this is quite uncertain, in part because eccentric waveforms for MBHB systems have only recently started to become available\cite[e.g.][]{PhysRevD.105.044035}. 

\subsection{Galaxy extraction}

The extraction of the GB population is the largest and most difficult part of the LISA global fit (at least for currently simulated datasets). Overall, our global fit algorithm performed very well. It was able to confidently find, characterize, and remove the detectable GB sources. It is more energy efficient than other pipelines, such as~\cite{Littenberg:2023xpl}, and we expect the difference in the efficiency between CPU and GPU algorithms to increase over time as discussed in previous sections.

One aspect of the sampler setup that should be further investigated in the future is the dynamic selection of dimensions within the temperature ladder. Currently, we use dynamic temperature adaptation to adapt the temperatures over the course of the run, but the number of walkers and temperatures for each sub-band is kept fixed and equal across the full frequency range. Ideally, we would place more resources where they are needed. For example, we should remove temperatures/walkers from edge samplers that do not need as large of a dynamic range due to the low number of templates needed to fit the data within those sub-bands. We would then use these resources in the center of the band, where the highest source counts and confusion exist, by adding temperatures/walkers. There are a few small planned changes to our algorithm to accomplish this goal. 

We also need to examine a larger variety of astrophysical models for the GB population, e.g., eccentric, triple, or accreting systems. The detached GBs are expected to make up the majority of the resolved catalog, so the fact that our algorithm performed so well means the largest piece of the fit is in place. Adding other astrophysical GB populations will complicate the fit, but probably only within a subset of the overall galaxy fitting problem. The two most difficult aspects of adding additional astrophysics to the GB population (in terms of data analysis) will be to create accurate, efficient template models and to understand how well those models can be matched by a basic detached GB template. If the match is high, it complicates the RJMCMC process by requiring efficient interchange between the model types. This is a topic for future work.

\subsection{Noise and instrument analysis}

For the current assumptions used to construct the LDC2A dataset, our instrumental noise and confusion signal PSD fitting algorithm performed well. It was able to fit both features and produce good posterior information.

The noise model assumed for the LDC2A analysis was purely stationary and Gaussian, including for the foreground. 
In reality, the foreground noise is cyclo-stationary. As the LISA constellation orbits the sun, the foreground confusion gains a directionality as the constellation moves toward and away from the Galactic center. The fit that we give is effectively an average of this time-dependence over the course of the observation. This not only affects our information about the Galaxy, it also affects the sensitivity used for each individual MBHB merger. The foreground noise locally in time to the MBHB merger may be above or below the average estimated from the full-year fit. This will affect the characterization of these signals~\cite{Digman:2022jmp}.

In the future, we will also have to include realistic orbits and instrumental effects in our global fit. This means using numerically generated orbits to construct the LISA response and include fitting for non-stationary noise from sensitivity drift, gaps, glitches, etc. in the global fit. 
Additionally, we will move to a more instrument-centric PSD parameterization to both input useful prior information into the global fit and to extract information about the instrument performance from it. 


\section{\label{sec:conclusion}Conclusion}

We have presented a new global fit algorithm capable of successfully analyzing the LDC2A training and hidden datasets from start to finish with no human intervention. We extracted and characterized all MBHB signals in both datasets without any false alarms. Our Galaxy sampler extracted $\sim10^4$ GBs with $>90\%$ match rate to the input population. Along with these astrophysical sources, we were able to fit the PSD and foreground confusion noise. 

The most important advancements in this pipeline are in the Galaxy sampler because of its large computational burden: i) GPU-accelerated sampling operations; ii) ensemble sampling and tempering; iii) and online fitting of RJMCMC distributions via refitting of samples and single-source MCMC runs against the current residual. The GPU-accelerated pipeline is more efficient in energy usage and cost than comparable CPU-based pipelines, and is expected to continue to improve into the future. The ensemble sampling allows for better mixing and better marginalization between source and noise modules across the global fit. The refitting and single-source MCMC pipelines used to generate the RJMCMC proposal distributions are very helpful to building up the galaxy solution. We expect these methods to extend to other sources as more advanced pipelines are constructed. We have also published our catalogs for both training and hidden LDC2A datasets~\cite{katz_training_data, katz_hidden_data}. Our new open-source algorithm and its catalog outputs now available to the public will be critical tools for encouraging access to and advancement of LISA data and astrophysical analysis across the global LISA community.

\section*{\label{sec:data_avail} Data/code availability}

The analysis codes (and the specific commits) used were \href{https://github.com/mikekatz04/LISAanalysistools/tree/9305e9963b1920d32fcb74fb341dd136063577c5}{LISA Analysis Tools}~\cite{lisaanalysistools_zenodo}, \href{https://github.com/mikekatz04/Eryn/tree/58347f5433ed9670b3c0791f5e395c5d00bea865}{Eryn}~\cite{eryn_zenodo}, \href{https://github.com/mikekatz04/BBHx/tree/646a4e268e37e4e1d5c87d0d94bdd4e8ce9b69e9}{BBHx}~\cite{bbhx_zenodo}, and \href{https://github.com/mikekatz04/GBGPU/tree/0a9e8429f64d4823c00cfe572c93c2f1da8181c0}{GBGPU}~\cite{gbgpu_zenodo}. We plan to clean up the code and add documentation and tutorials in the coming months. At this point, the full code will be released and made readily available to all. 

Our output catalogs are available for download on Zenodo. The training dataset catalogs~\cite{katz_training_data} can be found \href{https://zenodo.org/records/11000534}{here}. The hidden dataset catalogs~\cite{katz_hidden_data} can be found \href{https://zenodo.org/records/10989860}{here}. These data can be read-in and analyzed with \texttt{lisacattools}~\cite{lisacattools} and LISA Analysis Tools~\cite{lisaanalysistools_zenodo}.

\acknowledgements
The authors would like to thank Tyson Littenberg for helpful discussions in finishing this initial implementation. We would also like to thank Christian Chapman-Bird, Lorenzo Speri, Ollie Burke, Stanislav Babak, Sylvain Marsat, Robert Rosati, Alvin Chua, amongst others for helpful discussions along the way. N.Korsakova
acknowledges support from the CNES for the exploration of LISA science. N.Karnesis acknowledges the funding from the European Union’s Horizon 2020
research and innovation programme under the Marie Skłodowska-Curie grant agreement No 101065596. MLK would also like to thank Jelena Jelu\v{s}i\'c for her support while working on this paper. MLK also ackowledges the support of Kurrie and Boo for being good boys. This code made use of NumPy~\cite{Harris:2020xlr}, Matplotlib~\cite{Hunter:2007ouj}, SciPy~\cite{2020SciPy-NMeth}, and CuPy~\cite{Okuta2017CuPyA}.

\appendix

\section{GB Proposal algorithm}\label{sec:gb_proposals}

The GB proposal algorithm builds on the special GPU-based waveforms to handle the sampling and Likelihood computations in an efficient and light-weight manner. Algorithm~\ref{alg:gb1} describes the GB sampling implementation.

\begin{algorithm*}
    \caption{Galaxy sampling algorithm.}\label{alg:gb1}
    \begin{algorithmic}
        \State $j\gets(t, w, b)$ \Comment{All information contained for band $(t, w, b)$. Assume $b$ is odd.}
        \State $\mathbf{r}' \gets \mathbf{r}_{(0, w, b - 1)}^\text{even} + \mathbf{r}_{(0, w, b)}^\text{even} + \mathbf{r}_{(0, w, b + 1)}^\text{even}$  \Comment{Applies to frequencies contained in bands $b - 1$, $b$, and $b + 1$.} 
        \State $\mathbf{r}' \gets \mathbf{r}' - \sum_{\mathbf{\theta} \in \mathbf{\theta}_{j}}\mathbf{h}(\mathbf{\theta})$\Comment{Subtract current sources in band $b$, walker $w$, and temperature $t$.}
        \State $\mathbf{\theta}_{j, 0} \gets \mathbf{\theta}_{j}$
        \State $p\gets0$
        \While{$p < n_\text{prop}$}
            \State $k\gets\text{randint}(0, g_j)$ \Comment{$g_j$ is the number of GBs in band $j=(t, w, b)$.}
            \State $\mathbf{\theta}_k \gets \mathbf{\theta}_j[k]$
            \State $\mathbf{\theta}_k' \gets \text{Make Proposal on }\mathbf{\theta}$ \Comment{Includes accept/reject and Likelihood computations described in Equation~\ref{eq:swapsource}}
            \If{accepted}
                \State $\mathbf{\theta}_j[k] \gets \mathbf{\theta}_k'$
                \State $\mathbf{r}' \gets \mathbf{r}' + \mathbf{h}(\mathbf{\theta}_k)$ \Comment{Remove $\mathbf{h}(\mathbf{\theta}_k)$ from $r'$.}
                \State $\mathbf{r}' \gets \mathbf{r}' - \mathbf{h}(\mathbf{\theta}_k')$ \Comment{Add $\mathbf{h}(\mathbf{\theta}_k')$ to $r'$.}
            \EndIf
            \State $p\gets p + 1$
        \EndWhile
        \If{t = 0}
            \State $\mathbf{r}_{(0, w, b)} \gets \mathbf{r}_{(0, w, b)} + \sum_{\mathbf{\theta} \in \mathbf{\theta}_{j, 0}}\mathbf{h}(\mathbf{\theta})$ \Comment{Remove starting waveforms from residual in global memory.}
            \State $\mathbf{r}_{(0, w, b)} \gets \mathbf{r}_{(0, w, b)} - \sum_{\mathbf{\theta} \in \mathbf{\theta}_{j}}\mathbf{h}(\mathbf{\theta})$ \Comment{Add new group waveforms to residual in global memory.}
        \EndIf
    \end{algorithmic}
\end{algorithm*}

\begin{algorithm*}
    \caption{Galaxy Tempering algorithm.}\label{alg:gb2}
    \begin{algorithmic}
        \State $\mathbf{j} \gets(\mathbf{t}, \mathbf{p}, b)$ \Comment{$\mathbf{t}$ is the array over the temperature indices (range($n_t$)).} \State \Comment{$\mathbf{p}$ is the array of permuted walkers with len$(\mathbf{p})=n_t$.}
        \State Current Likelihood array: $\text{len}(\mathbf{\mathcal{L}}_c) = n_t$
        \State Swapped Likelihood array: $\text{len}(\mathbf{\mathcal{L}}_s) = n_t$
        \State $t\gets 0$
        \While{$t < n_t$}
            \State $w \gets \mathbf{p}[t]$ \Comment{$w$ is the walker index associated with temperature $t$.}
            \State $j \gets (t, w, b)$
             \State $\mathbf{r}' \gets \mathbf{r}_{(0, w, b-1)}  + \mathbf{r}_{(0, w, b+1)} -  \sum_{\mathbf{\theta} \in \mathbf{\theta}_{j}}\mathbf{h}(\mathbf{\theta})$

             \State $\mathbf{\mathcal{L}}_c[t] = \mathcal{L}(\mathbf{\theta}_{j})$
             \Comment{Store all current Likelihood values.}
             \State $t \gets t + 1$
        \EndWhile
        \State $t\gets n_t - 1$
        \While{$t > 0$}
            \State $w_t \gets \mathbf{p}[t]$
            \State $w_{t - 1} \gets \mathbf{p}[t - 1]$
            \State $\mathbf{r}' \gets \mathbf{r}_{(0, w_t, b-1)}  + \mathbf{r}_{(0, w_t, b+1)} -  \sum_{\mathbf{\theta} \in \mathbf{\theta}_{(t-1, w_{t-1}, b)}}\mathbf{h}(\mathbf{\theta})$
            \Comment{Residual in $w_t$ with sources from band $(t-1, w_{t-1}, b)$.}
            \State $\mathbf{\mathcal{L}}_p[t] \gets \mathcal{L}(\mathbf{\theta}_{(t-1, w_{t-1}, b)})$
             \Comment{Use $\mathbf{r}'$ in Likelihood computation.}

            \State $\mathbf{r}' \gets \mathbf{r}_{(0, w_{t-1}, b-1)}  + \mathbf{r}_{(0, w_{t-1}, b+1)} -  \sum_{\mathbf{\theta} \in \mathbf{\theta}_{(t, w_t, b)}}\mathbf{h}(\mathbf{\theta})$
            \Comment{Residual in $w_{t-1}$ with sources from band $(t, w_t, b)$.}
            
            \State $\mathbf{\mathcal{L}}_p[t-1] \gets \mathcal{L}(\mathbf{\theta}_{(t, w_t, b)})$
            \Comment{Use updated $\mathbf{r}'$ in Likelihood computation.}

            \State Accept or reject swap with Equation~\ref{eq:temp-final}.

            \If{swap accepted}
                \State $\theta' \gets \mathbf{\theta}_{(t, w_t, b)}$
                \State $\mathbf{\theta}_{(t, w_t, b)} \gets \mathbf{\theta}_{(t - 1, w_{t - 1}, b)}$
                \State $\mathbf{\theta}_{(t - 1, w_{t-1}, b)} \gets \theta'$
                \State $\mathbf{\mathcal{L}}_c[t] \gets \mathbf{\mathcal{L}}_p[t]$
                \State $\mathbf{\mathcal{L}}_c[t-1] \gets \mathbf{\mathcal{L}}_p[t-1]$

                \If{t = 1} \Comment{New cold-chain parameters ($t=1$, $t-1=0$).}
                     \State $\mathbf{r}_{(0, w_0, b)} \gets \mathbf{r}_{(0, w_0, b)}  + \sum_{\mathbf{\theta} \in \mathbf{\theta}_{(1, w_1, b)}}\mathbf{h}(\mathbf{\theta})$ \Comment{Remove old cold-chain sources from global residual.}
                     \State $\mathbf{r}_{(0, w_0, b)} \gets \mathbf{r}_{(0, w_0, b)}  - \sum_{\mathbf{\theta} \in \mathbf{\theta}_{(0, w_0, b)}}\mathbf{h}(\mathbf{\theta})$ \Comment{Add new cold-chain sources to global residual.}
    
                \EndIf
            \EndIf
            \State $t\gets t - 1$
        \EndWhile

    \end{algorithmic}
\end{algorithm*}


Due to complications with changing the dimensionality of the sorted and grouped GBs used in the proposal, the RJ proposals are currently run a bit differently from in-model Proposals. They are implemented in exactly the same way as in-model proposals, but the inputs are handled differently. For in-model proposals, the initial points and their associated information are organized and entered into the GPU kernel. It then runs multiple proposals for each binary and reads out the final information.

As discussed in Section~\ref{sec:rj_proposals} of the Appendix, RJ proposals are generated in a batched way across the whole band by proposing to remove all sources that are currently there and proposing to add a generated source for each array entry currently absent of a source. From an algorithmic perspective, we propose all points for the unused array entries upfront. This is true for all walkers, all temperatures, all unused array entries. We organize and group them with the sources that are in the sampler currently. This specialized grouping information is then input into the GPU kernel with two extra pieces of information: whether each binary is already in the sampler or not, as well as the PDF value for all sources (both existing and proposed) as determined from the generated distribution.

The GPU kernel then runs through all of these existing and proposed sources in randomized order within the same sub-band. We treat this like an in-model proposal simulating an RJ step by making the source of interest almost identical before and after the proposal, adjusting only its amplitude. If the source exists in the sampler, meaning it is a proposed removal, we set the initial amplitude to the true value and its proposed amplitude to a negligibly small number. If the source of interest is a proposed addition, we set the final amplitude to the true value and the initial amplitude to a negligibly small number. We accept or reject the proposed change after each individual proposed binary addition or removal. All sources and source parameters, other than for the specific source for which we propose a change, remain fixed throughout this sampling process.


\section{Setting up in-model GB proposals}

In-model proposals are MCMC proposals that do not change the dimensionality. In our case here, this is the underlying model count. As discussed in \cite{Karnesis:2023ras}, the regular stretch proposal does not scale well to reversible jump situations where the dimensionality varies. For this reason, we implement a ``group stretch proposal'' as detailed in \cite{Karnesis:2023ras}. The group stretch proposal is a generalization of the stretch proposal to reversible-jump situations with varying dimensionality and multiple intertwined maxima. The main idea of the group proposal is to set a stationary group of points from all leaves and walkers in the sampler from which to draw the auxiliary point needed in the stretch proposal. This point is drawn from a distribution of points in the stationary group that are ``close'' to the current point of interest from which the proposal originates. Here, ``close'' means as determined by some user-defined distance metric.

For the GB sampler, the initial frequency parameter is the most obvious parameter on which to base a distance metric to use to cluster points into nearby distributions. 
The GB group proposal begins by selecting the $n_\text{friends}$ points out of the previously stored stationary distribution that are closest in 
initial frequency parameter. It then draws uniformly from that group of nearby points to use in the stretch proposal.

In the limit where every walker contains every template and no template's posterior distribution interacts closely with another source, the group proposal will converge to the performance of the base stretch proposal. However, in realistic situations, the efficiency of the group proposal is highly situation dependent. 

We update the stationary distribution every $\sim30$ iterations. One downside to this setup is it does not work well when moving a template that appears in a low fraction of walkers (e.g., a source that is not real or a source that is marginally detectable). In this case, the moves are rarely accepted. While recognizing this shortcoming, based on empirical runs, we decided to use only this group proposal rather than dealing with the tuning and complications of implementing a covariance-based proposal. In the future, we plan to add a covariance proposal to help with these cases.

\section{RJ GB Proposal Basis}\label{sec:rj_proposals}

When running the RJ proposals, we do not use the typical ``birth-death'' model where one source is either added or subtracted for each RJ proposal. We set an array of $n_\text{bin, max}$ boolean values for each walker. True values in the array indicate a binary exists. False values indicate there is no binary at that array entry. The value of $n_\text{bin, max}$ is chosen to be sufficiently high so as not to affect the outcome of the experiment. During \textit{each} RJ proposal, we propose to switch every value of this array. We propose to remove all binaries that exist and add a binary for each value in our array that begins as ``False.'' We randomize the order of these additions and subtractions and run through them one-by-one using Algorithm~\ref{alg:gb1}. Theoretically speaking, this method is equally valid compared to the single birth-death setup usually used for nested RJ problems. Additionally, our RJ proposal distributions span the entire space, which we believe creates more flexibility and power in proposal generation.

\section{Parallel tempering in a residual-based analysis}\label{sec:parallel-temp}
We leverage parallel tempering in every module in our global fit sampler. Parallel tempering ``heats up'' the Likelihood surface by raising it to the power of the inverse temperature: $\mathcal{L}_\beta = \mathcal{L}^\beta$, with $\beta=1/T$. Samples are then probabilistically swapped between temperature chains (see below). This allows for the algorithm to explore the full Likelihood surface using the higher temperature chains to move between Likelihood maxima. As the sampling proceeds, the temperatures across the chains adapt to optimize swapping percentages according to the prescription given in~\cite{Vousden2016}. You can find more information about parallel tempering and its implementation in \texttt{Eryn} in~\cite{Karnesis:2023ras}.

In all of the sampling discussed below, we leverage the ensemble nature of \texttt{Eryn} which allows for many walkers per temperature. The ensemble setup creates better tempered mixing because the order of walkers within a temperature is randomly permuted before proposing temperature swaps. This means that, from one step to the next, sets of walkers for proposed swaps will always be newly randomly generated. 

The typical expression for the acceptance, $\alpha_{T,1}$, of temperature swaps between the parameters, $\vec\theta_n$ and $\vec\theta_m$, of walkers at temperatures $1/T_i=\beta_i$ and $1/T_j=\beta_j$ is given by~\cite{Vousden2016},
\begin{equation}\label{eq:temp-orig}
    \alpha_{T,1} = \text{min}\left[ 1, \left( \frac{\mathcal{L}_n}{\mathcal{L}_m} \right)^{\beta_j - \beta_i} \right],
\end{equation}
where ${\cal L}_k$ denotes the cold-chain Likelihood evaluated at the parameters $\vec\theta_k$.

In our global fit algorithm, each source type and the overall noise has its own temperature ladder setup, only sharing cold-chain information with each other. We use the ensemble setup within each source or noise module to marginalize over different instances of residuals (GBs plus MBHBs) and total noise PSDs (total means including the stochastic confusion noise) that come from the posterior distributions of other modules. For example, if we have $n_w$ walkers in each temperature, we may choose to have $n_w$ different realizations of the PSD or $n_w$ different residuals from the MBHB analysis in the GB part of the sampler (or vice-versa).

Having residual or PSDs attached to a specific walker complicates the swapping of samples between rungs on the temperature ladder. Equation~\ref{eq:temp-orig} assumes that the Likelihoods for the two states being swapped are equivalent. This is generally not the case if the two walkers have different residuals or PSDs. Suppose that we are proposing a swap between a walker, labelled $i$, in the chain at inverse temperature $\beta_I$, currently at parameter values $\vec{\theta}_n$, with a walker, $j$, at temperature $\beta_{II}$, at parameter values $\vec\theta_m$. We denote the cold-chain Likelihood for walker $i$, evaluated at $\vec{\theta}_n$ by ${\cal L}_{i,n}$. When deciding whether to swap the parameter values, $\vec\theta_n$ and $\vec\theta_m$, of the two walkers, we must account for the difference between the Likelihood functions associated with the walkers. The correct acceptance fraction, $\alpha_{T,2}$, is given by,
\begin{equation}\label{eq:temp-final}
    \alpha_{T,2} = \text{min}\left[ 1, \frac{(\mathcal{L}_{i,m})^{\beta_I}(\mathcal{L}_{j,n})^{\beta_{II}}}{(\mathcal{L}_{i,n})^{\beta_I}(\mathcal{L}_{j,m})^{\beta_{II}}} \right].
\end{equation}
Notice that in the case where the two walkers are sampling the same Likelihood  we have (\mbox{$\mathcal{L}_{i,n} = \mathcal{L}_{j,n}=\mathcal{L}_n$} and \mbox{$\mathcal{L}_{j,m}=\mathcal{L}_{i,m}=\mathcal{L}_m$}), Equation~\ref{eq:temp-final} reduces to Equation~\ref{eq:temp-orig} as expected. 

We use independent temperature ladders that adapt on their own
This type of issue has important ramifications for the efficiency of the sampler. It effectively means that for every swap proposed, two new Likelihoods need to be computed in addition to the original Likelihoods that enter the swapping operation. We will discuss how this applies differently to each module of the global fit in more detail below and in Algorithms~\ref{alg:mbh},~\ref{alg:gb1}, and~\ref{alg:gb2}. We use independent temperature ladders that adapt on their own for PSD, each different MBHB, and each sub-band in our GB sampler (see below). We communicate information between these ladders only through the cold chain ($T=1$).



\section{GB Tempering algorithm}

The tempering algorithm is very similar to the proposal algorithm with a few important changes to the information passed and how it is passed. This includes the ordering of the sub-bands to ensure permutation of walkers during temperature swaps. The tempering scheme is shown in Algorithm~\ref{alg:gb2}.

\section{GB Waveform Generation}\label{sec:gb_waveform}

Besides the source count uncertainty, the most difficult aspect of the GB algorithm is the large scale needed. With $\sim10^4$ binaries expected from $\sim1$ year of observation, even when using the most efficient waveform generators, the sheer number of waveforms needed throughout the MCMC run can be computationally difficult or prohibitive. To address this, we have developed an extremely efficient waveform generation pipeline.

Ultimately, we use functions based on \texttt{GBGPU}~\cite{gbgpu_zenodo} for generating GB waveforms efficiently. However, we use a special implementation that heavily leverages the usage of GPU Blocks and Shared Memory to make the waveform generation as efficient and light-weight as possible. For this task, we implement the waveforms in C/CUDA. We use a special FFT CUDA package~\cite{cufftdx} to perform FFTs of the slow-part of the waveform within GPU shared memory. Therefore, we never read out the waveforms to global memory, which ensures maximal efficiency for our computations. 

This waveform generation method can build the most common waveform in our analysis (256 frequency bins) at a rate of one waveform per $\sim200$ ns when properly batched. This is roughly $\sim1000$ times faster than those generated on a single CPU. From initial estimates, this translates to a power usage efficiency improvement of $\sim70$x.

\section{Algorithm for fitting proposal distributions}\label{sec:fitting}

While the other modules are running on their own GPUs, one extra process and one or two extra GPUs are used to search for and continuously improve upon the proposal distributions for the GBs. The cost of adding this GPU operation needs to be weighed against its helpful increase in efficiency of the overall process. We believe this step is incredibly helpful for getting the sampler to converge more rapidly.

There are two types of distribution fitting performed in the GB proposal fitting module. The first is the fitting of the distributions generated from searching for new binaries in each sub-band. The algorithm for this is a customized, batched, parallelized fixed-dimensional MCMC code based on \texttt{eryn}. It runs with $\sim10$ temperatures and $\sim100$ walkers per temperature using the stretch proposal to find the maximum Likelihood source that still exists in a given sub-band (meaning it is yet to be found and subtracted from the residual data). 

This algorithm runs simultaneous, independent MCMC runs for each sub-band ($\sim900$). The priors are modified so that the initial frequency spans its assigned sub-band only. Therefore, in this setup, the sources cannot move across sub-band edges. 

The starting points for each independent MCMC are generated from these modified priors. A user-defined threshold, $n_{\rm it,max}$, is set to determine the number of iterations required for maximum Likelihood convergence. This means the sampler must not find (within each specific sub-band) a higher Likelihood value for $n_{\rm it,max}$ consecutive iterations. For the results presented here we use $n_{\rm it,max}$=500.

Once this convergence criterion is reached, the best Likelihood point is taken for each sub-band. A new ``ball'' of points around this max Likelihood point is generated. The MCMCs are then relaunched with the same settings and stopping criterion. After this converges, we can be sure we have found the maximum Likelihood point available in each sub-band. We then do a test of each source against an optimal SNR limit of 7 to start. We eventually lower this to 5 over the course of the run (see Section~\ref{sec:search_to_pe} for more information). The MCMC is then run again for each source in each sub-band with an SNR above the limit. We generate 30 samples (after thinning by a factor of 25) for each of the 100 walkers giving us 3000 samples per source.

When all of these sample groups have been generated, we use \texttt{mpi4py}~\cite{mpi4py1, mpi4py2, mpi4py3, mpi4py4} to spread out this process to multiple processors which use \texttt{scikit-learn}~\cite{scikit-learn} to fit the GMM distributions. Ideally, this operation would be performed in batches on a GPU. This is a topic for future work. 


We also fit the points that have been in the large-scale sampler recently. This includes all sources in the cold chain that are appreciably detectable (optimal SNR$\gtrsim7$). We group these sources based on our cataloging operation described in Section~\ref{sec:cataloging}. After they have been grouped, we perform the same type of GMM fitting described above for the search proposal distributions. As mentioned previously, the purpose of this proposal is to help the mixing of the higher temperature chains by providing them with more direct proposals for binaries that are at higher SNR. 

In the future, we also plan to include GB population distributions fit with normalizing flows~\cite{Korsakova:2024sut}. We will investigate using these fitted distributions as our prior and/or proposal distribution functions.

\section{MBHB Proposal Algorithm}\label{sec:mbh_prop}

The MBHB proposal algorithm is given in Algorithm~\ref{alg:mbh}.

\begin{algorithm*}
\caption{MBHB Sampling Algorithm within the global fit.}\label{alg:mbh}
\begin{algorithmic}
    \State $j\gets0$
    \While{$j < n_\text{steps}$}
        \State $\vec{n}\gets$  permutation(range(0, $n_\text{mbhb}$)) \Comment{Randomly order the MBHBs each time.}
        \State $i \gets 0$
        \While{$i < \text{len}(\mathbf{n})$}
            \State $l \gets \mathbf{n}[i]$
            \State $\mathbf{h}_{(0,\mathbf{w}, l)} \gets f(\mathbf{\theta}_{(0, \mathbf{w}, l)})$ \Comment{$\mathbf{w}$ represents all walker indexes here. The ``0'' is the temperature index.}
            \State $\mathbf{r}_\mathbf{w}' \gets \mathbf{r}_\mathbf{w} + \mathbf{h}_{(0,\mathbf{w}, l)}$\Comment{Remove current cold chain waveforms from residuals ($\mathbf{d}_w - \mathbf{H}_w$).}
            \State $w_\text{ref} \gets \argmax_{\ \mathbf{\theta}_{(0, \mathbf{w}, l)}}\left[\mathcal{L}(\mathbf{\theta}_{(0, \mathbf{w}, l)})\right]$\Comment{Use maximum Likelihood point as reference waveform for Heterodyning.}
            \State $\mathbf{h}_\text{ref} \gets \mathbf{h}_{(0, w_\text{ref}, l)}$
            \State $p\gets0$
            \While{$p < n_\text{prop}$}
                \State Make proposal step for all $(\mathbf{t}, \mathbf{w}, l)$ \Comment{Use heterodyned Likelihood with residual $\mathbf{r}_{\mathbf{w}}'$ and reference waveform $\mathbf{h}_\text{ref}$.}
            \State $p\gets p+1$
            \EndWhile
            \State $\mathbf{h}_{(0,\mathbf{w}, l)}' \gets f(\mathbf{\theta}_{(0, \mathbf{w}, l)})$
            \State $\mathbf{r}_\mathbf{w} \gets \mathbf{r}_\mathbf{w}' - \mathbf{h}_{(0,\mathbf{w}, l)}'$ \Comment{Add the new cold chain waveforms back into the residual ($\mathbf{d}_w - \mathbf{H}_w$).}
            \State $i\gets i+1$
        \EndWhile
        \State FILE$\gets \mathbf{\theta_{(\mathbf{t}, \mathbf{w}, \mathbf{l})}}$
        \State HEAD PROCESS $\gets \mathbf{\theta_{(0, \mathbf{w}, \mathbf{l})}}$
        \State $\mathbf{r}_\mathbf{w} \gets \mathbf{r}_{\mathbf{w}, \text{new}}$ \Comment{Residual update from HEAD PROCESS.}
        \State $j\gets j + 1$
        \EndWhile
    \end{algorithmic}
\end{algorithm*}

\bibliography{globalfit}

\end{document}